\documentclass[preprint2]{aastex}

\shorttitle{Methanol \& OH Maser Survey}
\shortauthors{Cotton et al.}

\begin{document}

\title{A Large--Scale Spectroscopic Survey of Methanol and OH Line
Emission from the Galactic Center: Observations and Data}



\author{ W.~D.~Cotton\altaffilmark{1}}
\affil{National Radio Astronomy Observatory, 520 Edgemont Road,
Charlottesville, VA 22903}

\author{F.~Yusef-Zadeh} 
\affil{CIERA, Department of Physics and Astronomy, Northwestern University, Evanston, IL 60208}



\altaffiltext{1}{The National Radio Astronomy Observatory is a
facility of the National Science Foundation operated by Associated
Universities, Inc.} 


\begin{abstract}
Class I methanol masers are collisionally pumped and are 
generally correlated with outflows in star forming sites in the Galaxy. 
Using the VLA in its A-array configuration, we present a spectral line survey to identify methanol
$J=4_{-1}\rightarrow3_0E$ emission at 36.169~GHz.  
Over 900 pointings were used to cover a region 66$'$ $\times$ 13$'$ along
the inner Galactic plane. 
A shallow survey of OH at 1612, 1665, 1667 and 1720 MHz was also
carried out over the area covered by our methanol survey.  
  We provide a catalog of 2240 methanol masers with narrow line-widths
of $\sim1$ km s$^{-1}$, spatial resolution of $\sim0.14''\times0.05''$ and 
RMS noise $\sim20$ mJy beam$^{-1}$ per channel. 
Lower limits on the brightness temperature range from 27,000 K to
10,000,000 K showing the emission is of non-thermal origin. 
We also provide a list of  
 23 OH (1612), 14 OH (1665), 5 OH (1667)
and 5 OH(1720 MHz) masers.   
The origin of such a large number of methanol masers is not clear. 
Many methanol masers appear to be associated with infrared dark clouds, 
though  it 
appears unlikely that these masers trace early phase of star formation in the Galactic
center.    
%
\end{abstract}





\section{Introduction}
The inner few hundred pc of the Galactic center differs from the rest
of the Galaxy in its ISM properties. 
This region is occupied by a large concentration of warm molecular gas
with high column density, high velocity dispersion, high gas
temperature and high cosmic ray flux
\cite{bally88,huettemeister93,dahmen97,oka98,tsuboi99,oka05,larosa05,voronkov06,
jones12,yusef-zadeh13}.
What is  unusual about this region is that the gas characteristics 
resemble those of  star forming molecular cores  and yet few 
sites of star formation are recognized in the inner degree of the 
Galactic  center. 
To investigate the origin of hot molecular gas, 
we surveyed the Galactic center region using the 
36.2 GHz methanol line emission. 
The 36.2 GHz methanol  maser  is generally associated with
collisionally excited outflow sources \citep[eg]{voronkov06}. 

{ The lower resolution C configuration  survey of the region $-0.6^\circ
< l <\ +0.5^\circ$ and $-0.1^\circ<b<+0.1^\circ$ by Yusef-Zadeh et al. (2013)
with the Very Large Array}
found 356 methanol maser candidates { at 36 GHz} with a spatial and spectral
resolutions of  1.8$''\times0.7''$ and  16.6 km s$^{-1}$,
respectively, with RMS noise $\approx$25 mJy beam$^{-1}$  \citet{Paper1}. 
These maser candidates were identified as part of continuum
observations using  the Ka band at 35 GHz. 
This { maser transition} was imaged using 41 channels of continuum
data. 
The source list was divided into strong and weak maser candidates with
line fluxes greater than or less than 10 Jy km  s$^{-1}$
($T_b\sim$446K for unresolved sources), respectively.    
This threshold was selected because the gas temperature of molecular clouds 
 in the Galactic center is
less than 446K, thus many methanol sources  { may be} tracing thermal
emission  with broad line-widths.  

{ We report the presence of 2240 methanol { and 47 OH} masers
 distributed toward { the region surveyed by Yusef-Zadeh et al. (2013)}.  
The high abundance of methanol traced by quasi-thermal and maser emission}
is unlikely to be produced by gas-phase reactions, 
thus a different heating mechanism involving grain-surface chemistry
is needed to evaporate methanol from dust grains \cite{mehringer97}. 
One possibility for the source of heating dust grains in the central
molecular zone (CMZ) of the Galactic center is the interaction of
cosmic rays and molecular clouds \citep{yusef-zadeh07}. 
In this picture, the abundance of methanol can be enhanced by induced
photodesorption by cosmic rays as they travel through a molecular core
\citep{roberts07,yusef-zadeh13}. 
Another byproduct of such an interaction is the increase in the 
abundance of OH, also known to be collisionally pumped and are
produced at the interaction site when a supernova shock runs into a
molecular cloud \citep{frail94,wardle99}.  
{ One of the signature of collisional pumping is that only the OH(1720
MHz) line is a maser  whereas 1665/1665, { 1612} MHz emission are
thermal and are generally detected in absorption (see Hewitt et
al. (2008) 
and references therein). 
The present work presents the { methanol and OH } data from this
survey and future publications will explore the correlation of
methanol masers with other probes of star formation activity such as
radiatively excited methanol and water masers in detail.}

\section{Observations and Data Reductions}
\subsection{Methanol Observations using the Ka Band}
The maser observations were centered on the rest frequency of the
collisionally excited class I methanol $J=4_{-1}\rightarrow3_0E$
masers ({ adopted rest frequency} 36.169265 GHz).  
This sub-band had 1024 $\times$ 62.5 kHz channels giving a velocity
resolution of $\approx$ 1.0 km $s^{-1}$ after Hanning smoothing. 
The observations also contained eleven 128 MHz sub-bands with 2 MHz
resolution between 35.3 and 36.5 GHz for continuum calibration.
{
The region of galactic longitude -0.6$^\circ$ to +0.5$^\circ$ and galactic latitude -0.1$^\circ$
to +0.1 $^\circ$ was covered in 900 pointings.
}

\begin{figure*}
\includegraphics[width=2.5in,angle=-90]{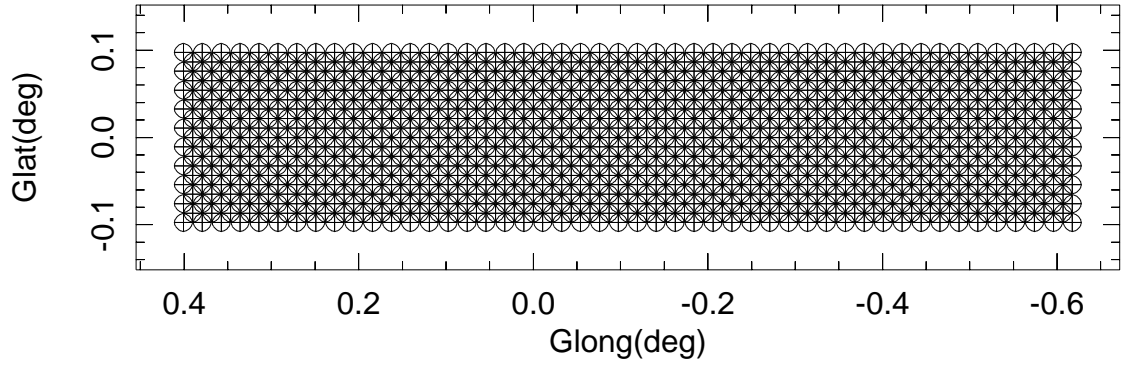}
\caption{Pointing centers of the Ka band observations and circles
showing the half power radius.
}
\label{KaPointfig}
\end{figure*}

These observations were made in five sessions between 2014 May 2 and
2014 { May} 22.
Each session included a 19 $\times$ 19 pointing hexagonal grid with
0.65$' $ separation of the centers; { the antenna FWHM is 1.25$'$.
The rasters were parallel to the galactic plane.  }
Each pointing was allocated 40 seconds of which 32 were typically on
source integrations.
Every 5th observation was of the phase reference calibrator,
J1744-3116, with pointing offsets measured hourly.
3C286 was used as the 
{ primary flux
}
and band-pass calibrator.
We did not apply Doppler corrections during our observations.
The array was in the ``A'' configuration giving a maximum baseline of
35 km; due to the southern declination of the target fields, the
synthesized beam is elongated north--south, typically 0.14$''$ $\times$ 0.05$''$.
Antennas detected Right-- and Left--hand circular polarization and all
four correlation products were recorded.
The 36 GHz pointing centers and half power radius is shown in Figure
\ref{KaPointfig}.


\subsubsection{Calibration\label{CH3OHCal}}
Calibration and imaging used the Obit package
\citep{Obit}\footnote{http://www.cv.nrao.edu/$\sim$bcotton/Obit.html}.
Calibration and editing was done on each session independently and
consisted of the following steps: 
\begin{enumerate}
\item The correlated data were only crudely calibrated in group delay
leaving strong derivatives of phase with frequency.
Residual group delay offsets were determined from the continuum
observations of 3C286 and J1744-3116 and applied to all data. 
\item Residual variation of gain and phase with frequency was
corrected by band-pass calibration using 3C286.
\item 
{ Flux density
}
calibration was based on a standard spectrum of
3C286 which was used to determine the spectrum of the astrometric
calibrator, J1744-3116.
Observations of J1744-3116 were then used to 
{ calibrate the amplitudes and phases of} all data.
\end{enumerate}
The spectroscopic data also had the following corrections:
\begin{enumerate}
\item Doppler corrections for the earth's motion were applied for each
pointing.
\item Data were Hanning smoothed to reduce the Gibbs ringing of the
stronger masers.  
\end{enumerate}
All data-sets were then subjected to a baseline dependent time
averaging to reduce the data volume.
This averaging was subject to the constraint that the amplitudes were
reduced by no more than 1\% to a radius of 0.6$'$ and averaging was no
more than 10 seconds.

\subsubsection{Methanol Imaging}
The 315 channels covering the velocity range 195 to -120 km $s^{-1}$ were
imaged with a velocity resolution of 1.0 km $s^{-1}$.
{ The masers generally have a  line width 
no larger than the spectral resolution of} 1 km$ s^{-1}$ so are typically seen
in a single channel.
Each pointing was imaged to a radius of 0.82 arcminutes from the pointing
with a grid spacing of $\approx$ 0.015$''$ and
was CLEANed to 300 mJy or a maximum of 300 components.
Images are typically 8000 $\times$ 8000 $\times$ 315 cells.
The resolution is approximately  0.14$''$ $\times$ 0.05$''$.
Due to the limited {\it uv} coverage and very large number of pixels in the
images, sidelobe levels are high and autoboxing was used to guide the
CLEAN.  
The RMS noise in channel images without significant emission is
$\approx$ 20 mJy beam$^{-1}$.
Each of the 900 single snapshot pointings was imaged independently and
the results combined in a linear mosaic. 

\subsubsection{Spectral cube mosaic}
Overlapping image cubes were combined into a ``mosaic'' for each
pointing using a grid of 6000$^2$ with a pixel size of 0.015$''$.
The images were combined channel--by--channel by multiplying each
channel image by the primary antenna pattern.
Grids of the image times the primary beam and the primary beam squared
were interpolated and summed into the combined grid.
When all overlapping pointing cubes were accumulated, the combined
image was normalized by dividing the sum of the weighted images times
antenna pattern by the sum of the antenna pattern squared.
This results in an optimally weighted primary beam corrected image
using all overlapping pointings. 

\subsubsection{Source finding\label{Finding}}
The relatively poor {\it uv} coverage of the single snapshot images results
in a quite strong variation of the quality of the images from pointing
to pointing depending on  what emission is in the field.   
There are vastly more independent resolution elements in the image
than independent {\it uv} samples in the data; this complicates the
detection of maser components. 
For source detection the ``mosaiced'' image cubes were collapsed to a
single plane by summing the ``significant'' pixels in each plane.
These were pixels in each image plane which were
in excess of eight times the off--source RMS and more than 0.25 times the
brightest pixel in the channel image.
This collapsed image was then searched for sources using Obit task
FndSou with a minimum acceptable flux density of 200 mJy beam$^{-1}$.
Candidate masers were then selected before they were fitted by 2-D Gaussians. 
Note, this procedure will select continuum as well as maser sources.

Candidate masers were then tested using the spectrum extracted at the
appropriate location from the combined spectral cube.
The average flux density in each pixel's spectrum was subtracted to
remove continuum emission.
Spectral features were accepted if the channel with the highest flux
density had a value in excess of 8 times the RMS measured within 50
pixels of the component.
This removes continuum sources as well as the bulk of the imaging
artifacts marked as candidate masers.
{ The sparse uv-coverage and very large number of image pixels of the
observations lead to a high level of imaging artifacts in regions with
significant emission. 
In order to avoid including many of these artifacts as spurious
sources, relatively conservative criteria are needed for source
selection. }
{ The likelihood of artifacts being included in the maser list
depends strongly on the number and strength of true maser emission in
the antenna beam. 
However, note that in areas where deeper, comparable resolution images
are available (Sjouwerman et al. 2010), the masers in
Table \ref{KaMaserCatalog} are confirmed.} 
The maser velocity and width were determined from a moment
analysis around the channel with the peak flux density.

Errors in positions given in Tables {\ref{ContKaCat}, and \ref{KaMaserCatalog}}
are fitting errors accounting for the 
effects of correlated pixel noise \citep{con97} and do not include
calibration errors.
Astrometric calibration errors are dominated by the transfer of
phases from the calibrator whose position is known to milliarcsecond
accuracy to the target data.
The calibrator, J1744-3116, is approximately 2$^\circ$ away on the sky
and was observed every 200 seconds and phases were interpolated to the
times of the target data.
Systematic errors should be under 1/3 of the sythesized beam which is
of the order of 30 milliarcseconds.

Errors in peak flux densities are derived from the fitted Gaussians in
the source finding step accounting for the effects of correlated
noise but do not include calibration errors.
Systematic flux density errors are dominated by 1) the transfer
of the flux density of the primary calibrator (3C286) to the phase
reference calibrator, 2) uncorrected atmospheric attenuation and
3) decorrelation due to atmospheric phase noise.
The primary calibrator was generally observed at a much higher
elevation than the phase reference calibrator which was at nearly the
same, low, elevation as the target fields.
Target fields with strong emission were phase self calibrated which
should effectively eliminate atmospheric decorrelation for these
pointings.
Systematic flux density errors should be no more than 30\%.

The final catalog was derived by combining the source lists from each
mosaic image and removing redundant entries from the overlap regions.



\subsubsection{Continuum Imaging}
The data from each pointing were imaged with a field of view of 1$'$
radius and CLEANed to a flux density of 0.5 mJy  beam$^{-1}$ or a
maximum of 300 components.
Pointings containing Sgr A* were phase self calibrated.
Overlapping pointings were combined on a ``mosaic'' at the position of
each pointing center on a 8000$\times$8000 grid of 0.012$''$ cells.
These images were searched for sources above 2 mJy beam$^{-1}$ or 8
times the RMS within 200 pixels.
RMS noise levels are typically 0.25 mJy beam$^{-1}$.

\subsection{Hydroxyl  Observations Using L Band }
\begin{figure}
\includegraphics[width=4.0in,angle=-90]{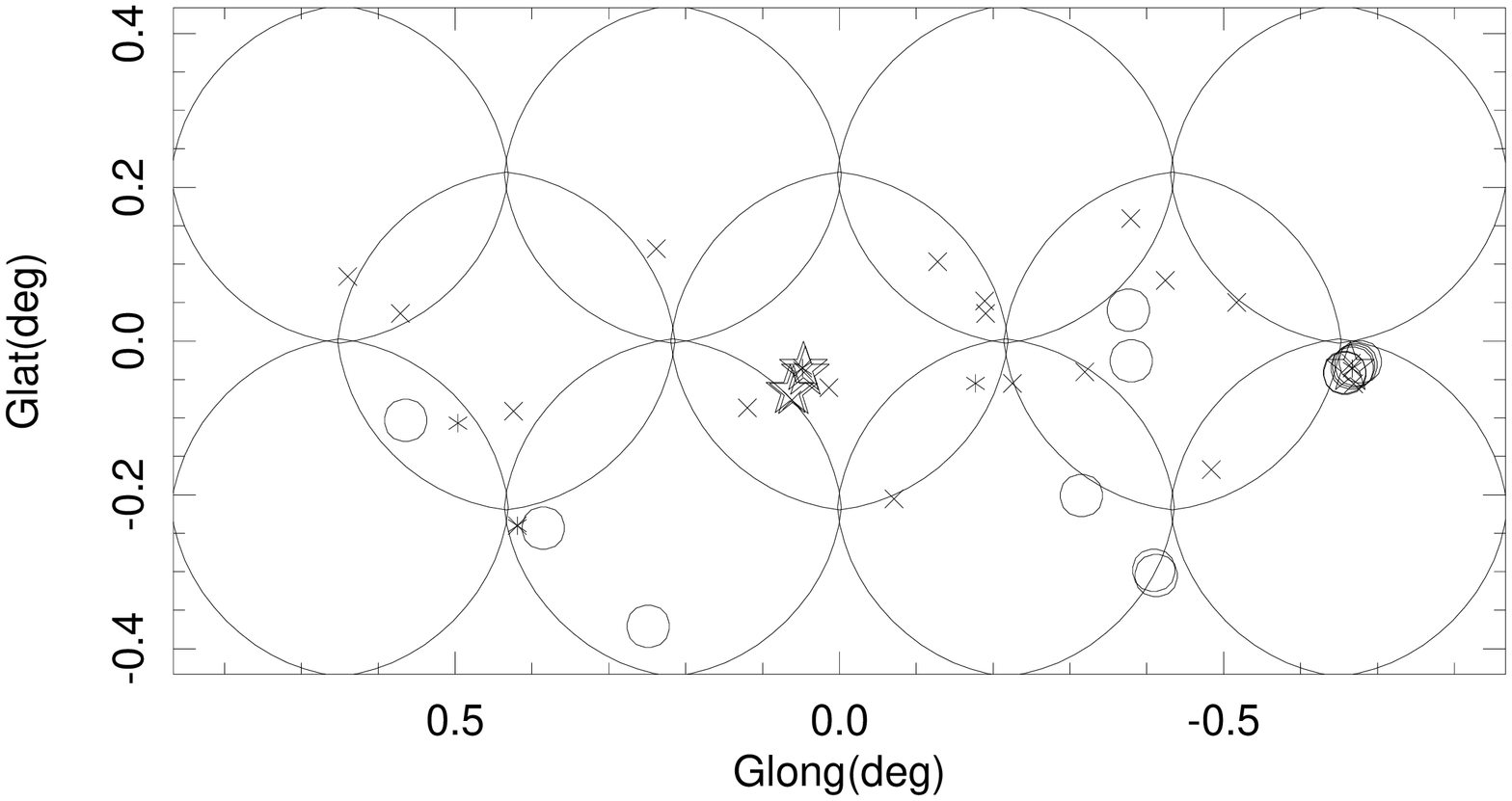}
\caption{Pointing centers of the L Band observations and locations of
OH masers; large circles show the half power radius.
``$\star$'' = 1720 MHz, ``*'' = 1667 MHz,
``$\circ$'' = 1665 MHz,  ``$\times$'' = 1612 MHz.
}
\label{LPointfig}
\end{figure}

The observations were 
{ made with spectral windows }
centered on the rest frequency of the OH lines at 1612,
1665, 1667 and 1720 MHz and  were made on May 31, 2014 using the VLA in ``A''
configuration.
This gives  a typical resolution of $\approx$ 3.5$''$ $\times$
1.2$''$ elongated north-south due to the declination of the region imaged.
The region was covered by a mosaic of 11 overlapping pointings
centered on positions given in Figure \ref{LPointfig}. 
{ The antenna beam size is $26'$.}
Each spectral window had 1024 $\times$ 3.9 kHz channels
giving a velocity resolution of $\approx$ 0.73 km $s^{-1}$.
The velocity range covered was $\approx$ -420 to +235 km $s^{-1}$.
The { adopted} rest frequencies were 1.612231, 1.66540184, 1.667359
and 1.72053 GHz. 
Each pointing was observed for 2 minutes in three scans separated by
about 45 minutes.
{ Flux density calibration used 3C286 and J1744-3116 was used as
  the phase reference calibrator.
}
Calibration followed the procedure given in Section \ref{CH3OHCal}
except that Hanning smoothing was not required.

\subsubsection{L Band Imaging}
The L Band data were imaged using faceting to cover the primary
beam.  
Continuum imaging was done on each pointing using all four sub-bands
out to a radius of 30$'$ and CLEANing 300 iterations.
Up to 3 iterations of phase self calibration was used for fields with
a peak brightness greater than 50 mJy beam$^{-1}$.
In the quieter fields, the off source RMS in the continuum images was
$\approx$1 mJy beam$^{-1}$ although with much variation and many
artifacts from the strong, almost completely resolved emission

Spectral line channels were imaged to a radius of 13.6$'$ and CLEANed up
to 20 iterations or a minimum residual of 1 Jy.
Typical RMSes in line free channels are $\approx$50 mJy beam$^{-1}$.
AutoBoxing was used to guide the CLEAN.
Combination of overlapping images was performed as for the 36 GHz
data but using a grid of 4800$\times$4800 pixels with a size of 0.3$5''$ each.
The OH emission spots are generally unresolved so only the lower limit
to the brightness temperature can be estimated.

As for the 36 GHz observations, errors in positions given in Tables \ref{ContLCat},
\ref{OH1612MaserCatalog},  \ref{OH1665MaserCatalog},
\ref{OH1667Spectfig} and  \ref{OH1720MaserCatalog} 
are fitting errors accounting for the 
effects of correlated pixel noise and do not include calibration errors.
Systematic astrometric calibration errors are dominated by the transfer of
phases from the calibrator whose position is known to milliarcsecond
accuracy to the target data.
The calibrator, J1744-3116, is approximately 2$^\circ$ away on the sky
and was observed every approximately 12 minutes and phases were
interpolated to the times of the target data.
Systematic errors should be under 1/10 of the sythesized beam which is
of the order of 200 milliarcseconds.

Errors in peak flux densities are derived from the fitted Gaussians in
the source finding step accounting for the effects of correlated
noise but do not include calibration errors.
Systematic flux density errors are dominated by 1) the transfer
of the flux density of the primary calibrator (3C286) to the phase
reference calibrator and 2) decorrelation due to atmospheric phase noise.
Target fields with strong emission were phase self calibrated which
should effectively eliminate atmospheric decorrelation for these
pointings.
Systematic flux density errors should be no more than 10\%.

\section{Results \& Discussion}
To distinguish between methanol masers and { the broad line methanol sources
of Yusef-Zadeh et al. 2013,}
we carried out follow-up  methanol observations using the A-array of
the Very Large Array (VLA)  
with a spatial and Hanning smoothed spectral resolutions of 
0.14$''\times0.05''$ and $\sim$ 1 km s$^{-1}$,  respectively, with RMS noise $\sim20$ mJy beam$^{-1}$. 
 We found a  total of 2240 methanol masers. 
 In addition, we carried out OH observations of the same region. 
Here we give details of these  new  observations and present a catalog
of spectral line emission from methanol (36.1 GHz), OH (1612 MHz), OH
(1665 MHz), OH (1667 MHz) and OH (1720 MHz) masers to examine the
spectral and spatial correlations of these masers.       
There are two small regions that have already been studied in earlier
spectroscopic measurements at the transition of 36.2 GHz methanol. 
One is the inner 3$'\times3'$ of Sgr A where the 20 and 50 km $^{-1}$
clouds are located 
\cite{sjouwerman10}. 
Another region that has also been studied at 36.2 GHz is G0.253+0.01
or the "Brick cloud" \citep{mills15}.  
A more detailed analysis of { the masers presented here}
and their  correlations will be given elsewhere.

\subsection{36.2 GHz Continuum and CH$_3$OH Masers}
\begin{figure*}
\centerline{
\includegraphics[width=2.5in,angle=-90]{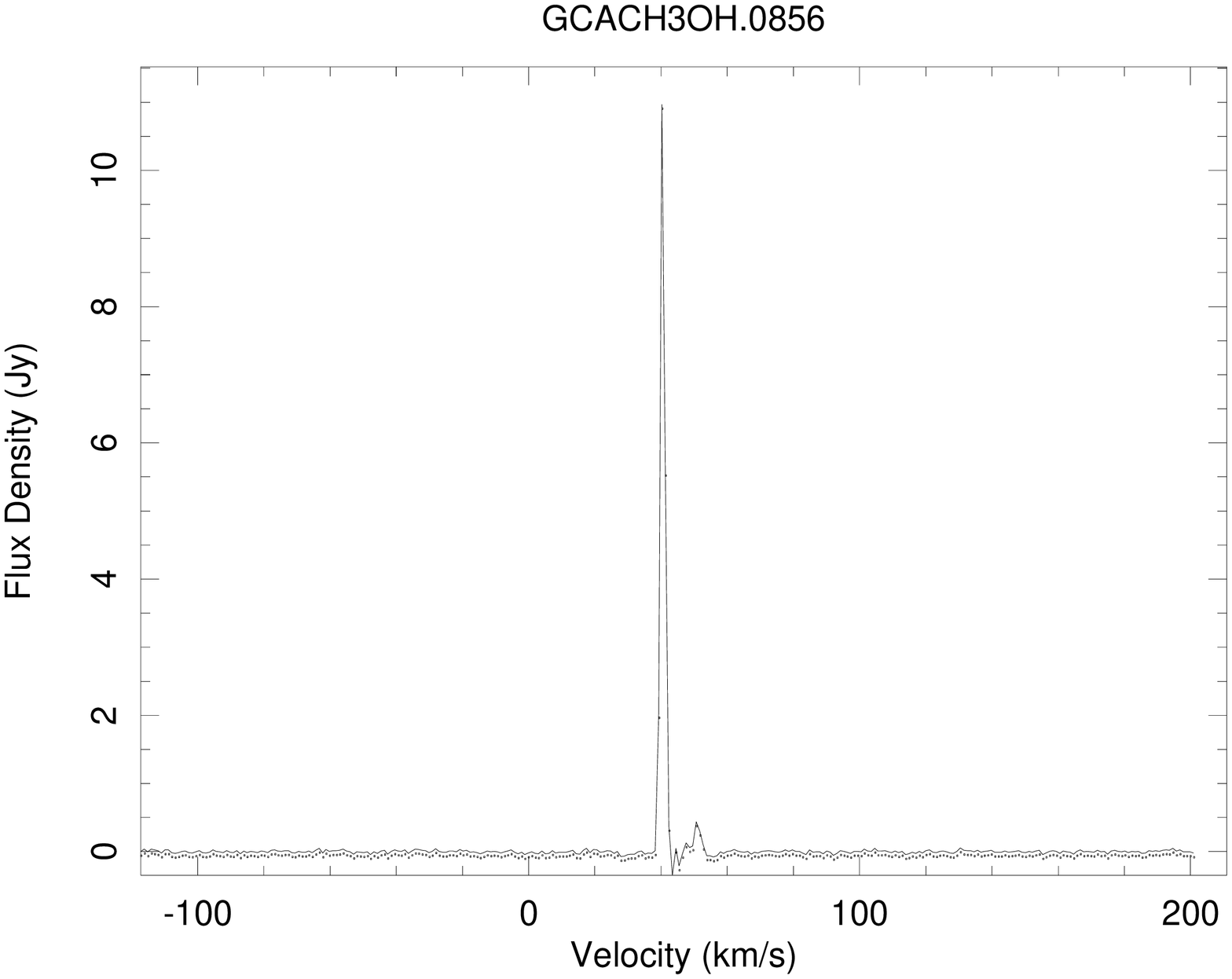}
\includegraphics[width=2.5in,angle=-90]{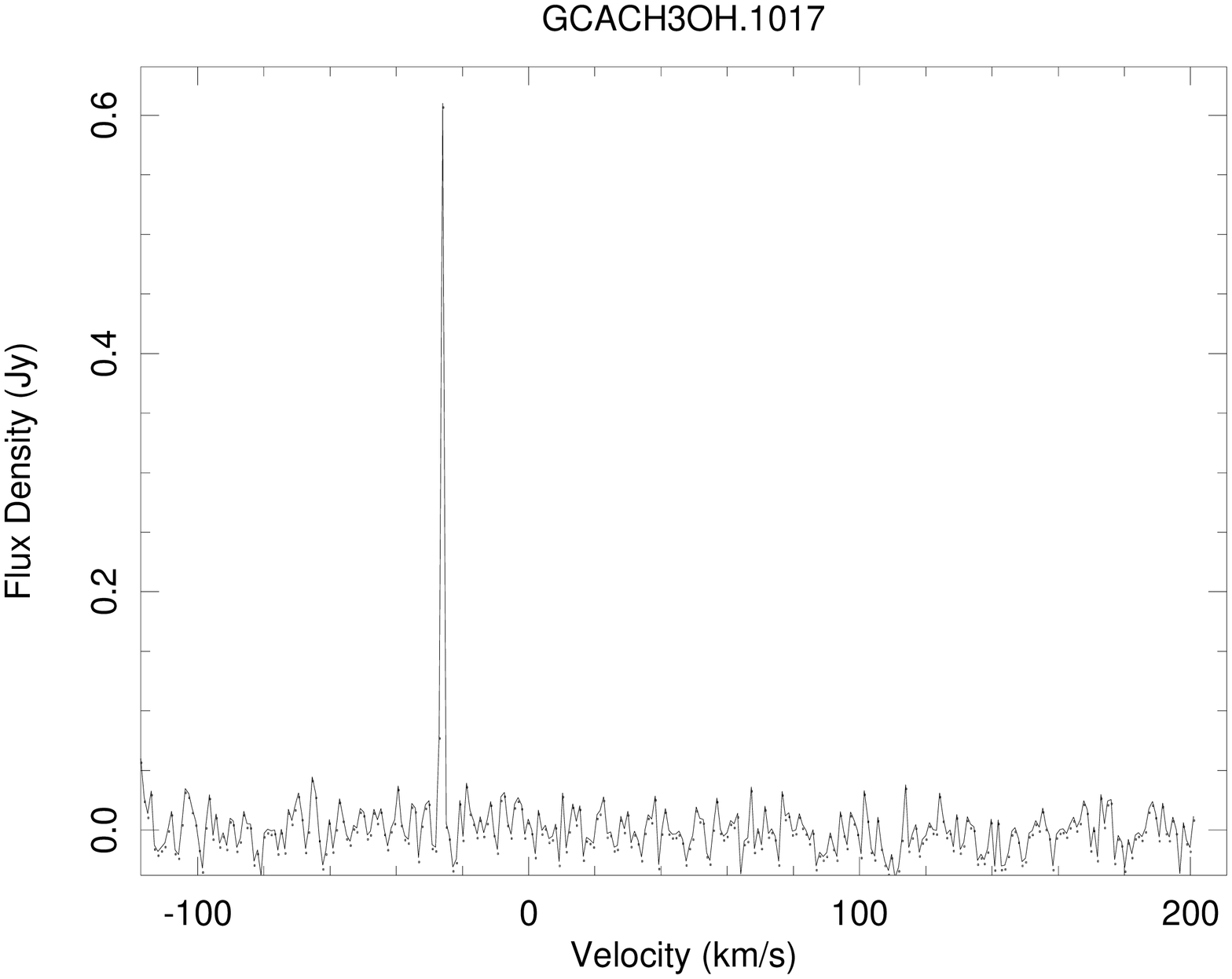}
}
\centerline{
\includegraphics[width=2.5in,angle=-90]{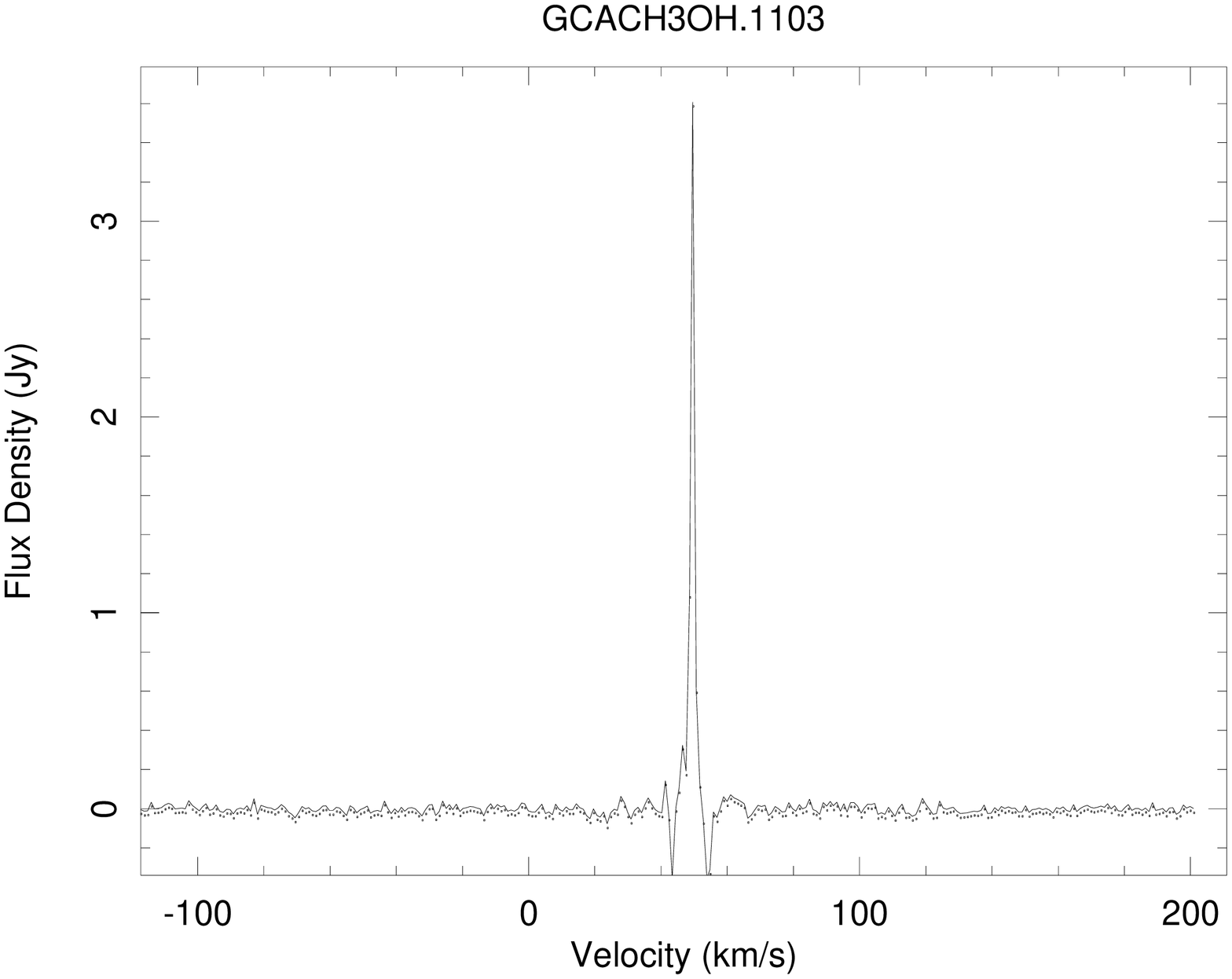}
\includegraphics[width=2.5in,angle=-90]{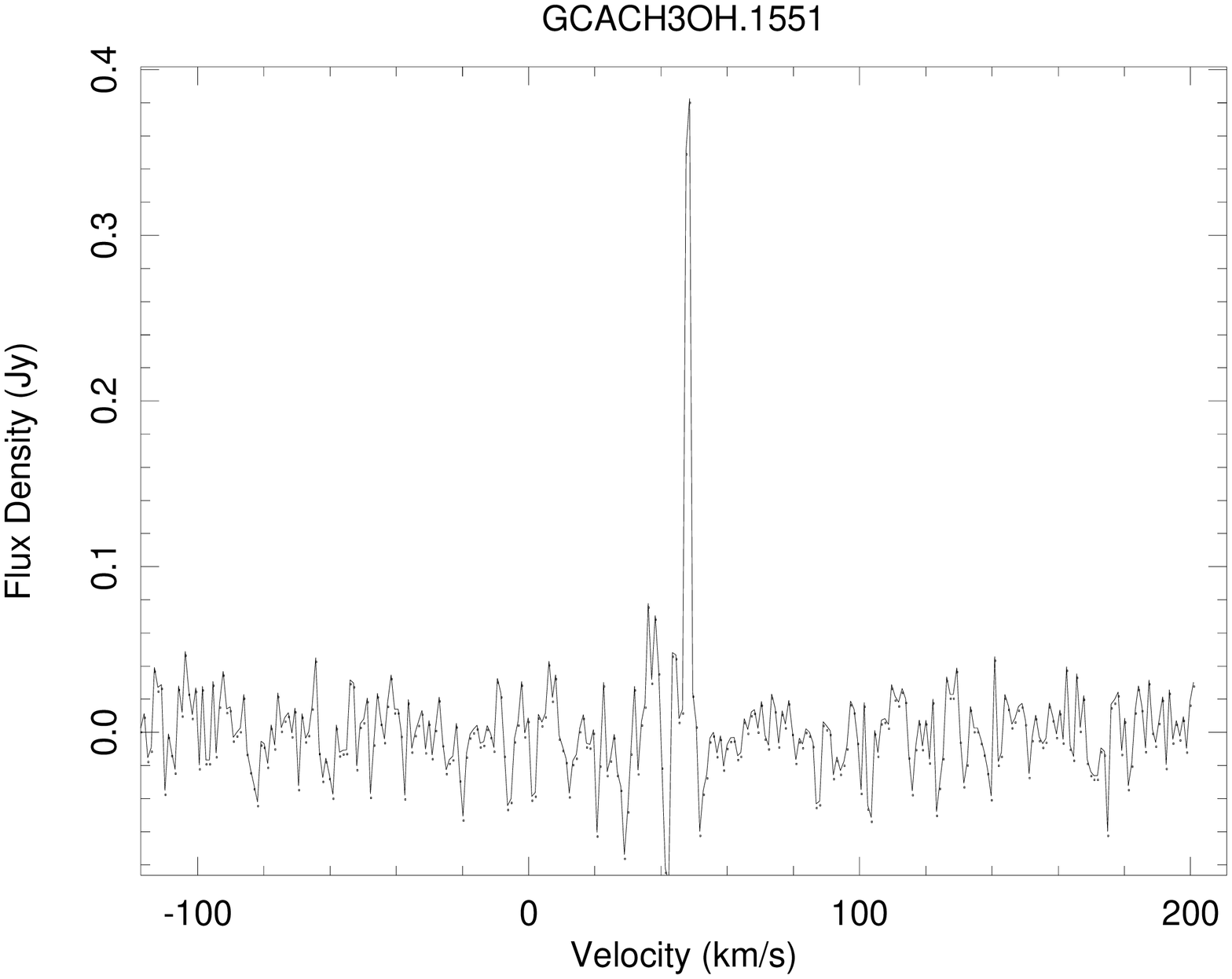}
}
\centerline{
\includegraphics[width=2.5in,angle=-90]{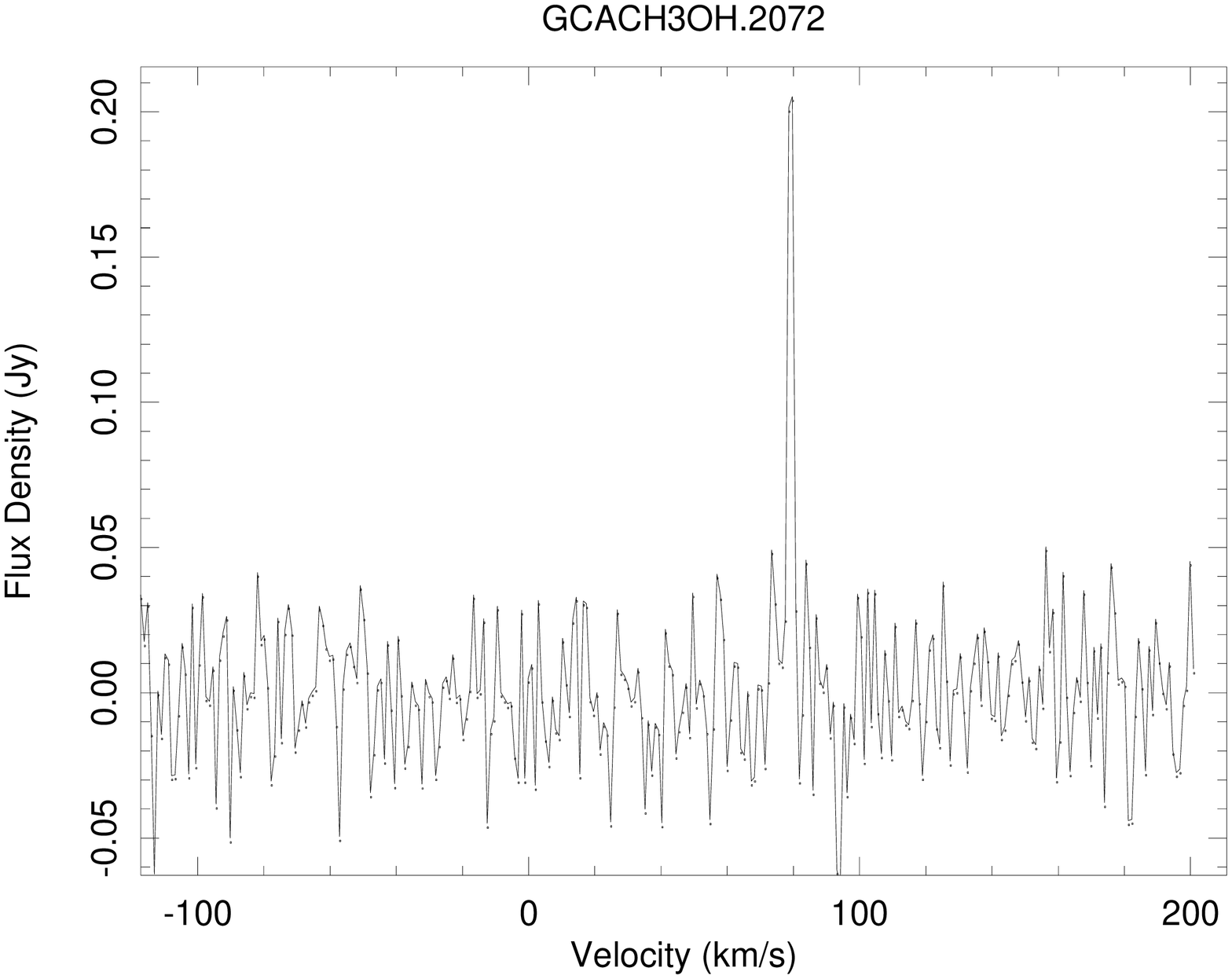}
\includegraphics[width=2.5in,angle=-90]{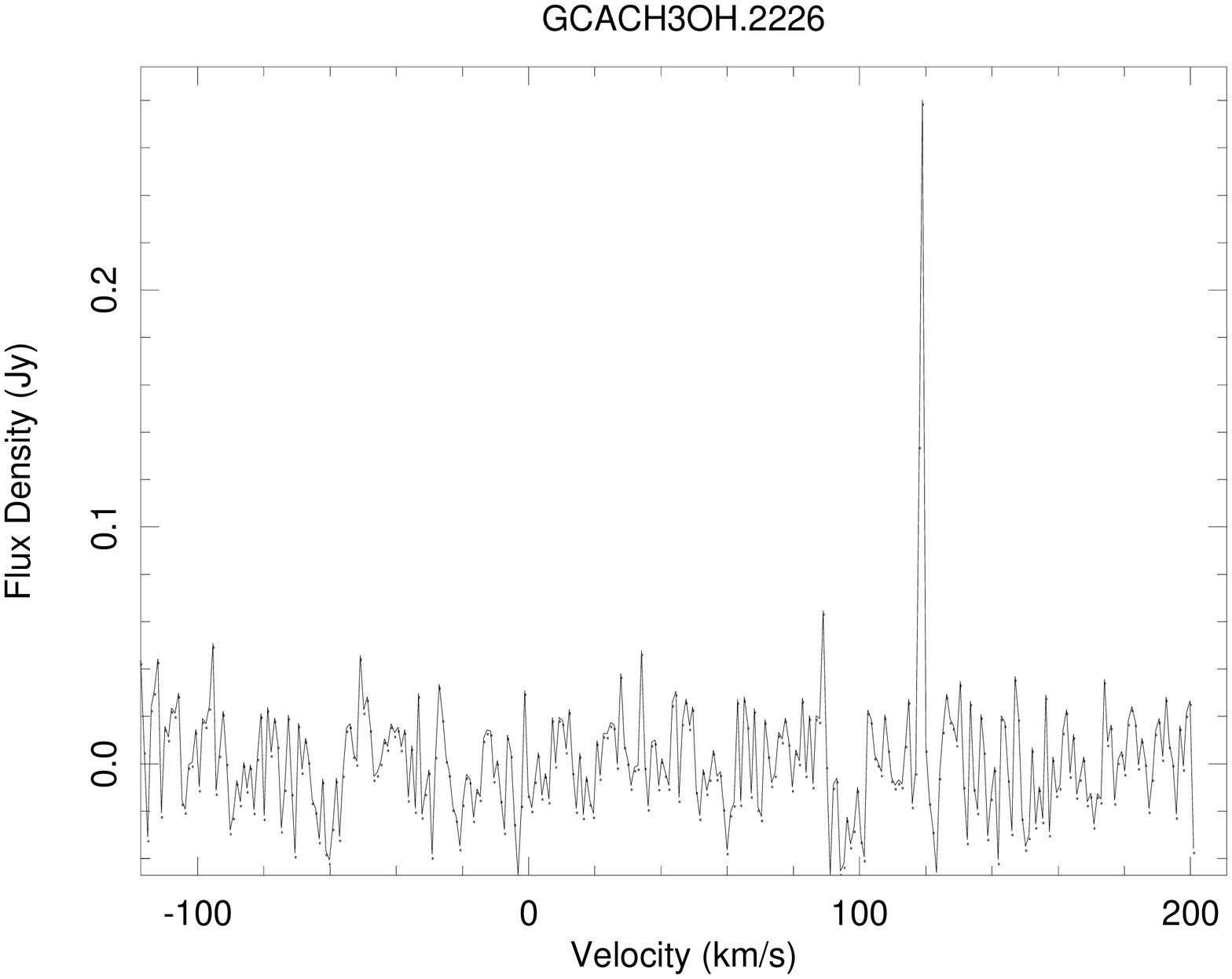}
}
\caption {Sample methanol maser spectra,
the source { name} is given above each plot.}
\label{CH3OHSpectfig}
\end{figure*}

The list of continuum sources is shown in Table \ref{ContKaCat}
listing 7 columns indicating the source { name}, celestial coordinates
and the flux density with their corresponding errors.  
A sample of  methanol sources is listed in Table \ref{KaMaserCatalog} with 11
columns giving:  the source  { name}, RA, Dec with errors, Galactic longitude
and latitude, the line flux with error, the center velocity, the
line-width and the brightness temperature.  
Because the sources are only marginally resolved in velocity and
generally unresolved spatially, only the lower limit to the brightness
temperature can be obtained; these are given in the last column in 
Table \ref{KaMaserCatalog}. 
The methanol sources in Table
\ref{KaMaserCatalog} are widespread with very narrow line-widths. 
The large lower bounds on the brightness temperature (27,000 K to
10,000,000 K) given in Table \ref{KaMaserCatalog}
support the interpretation of maser emission.  
A sample of spectra is shown in Figure \ref{CH3OHSpectfig}.

\begin{figure*}
 \includegraphics[height=2.1in]{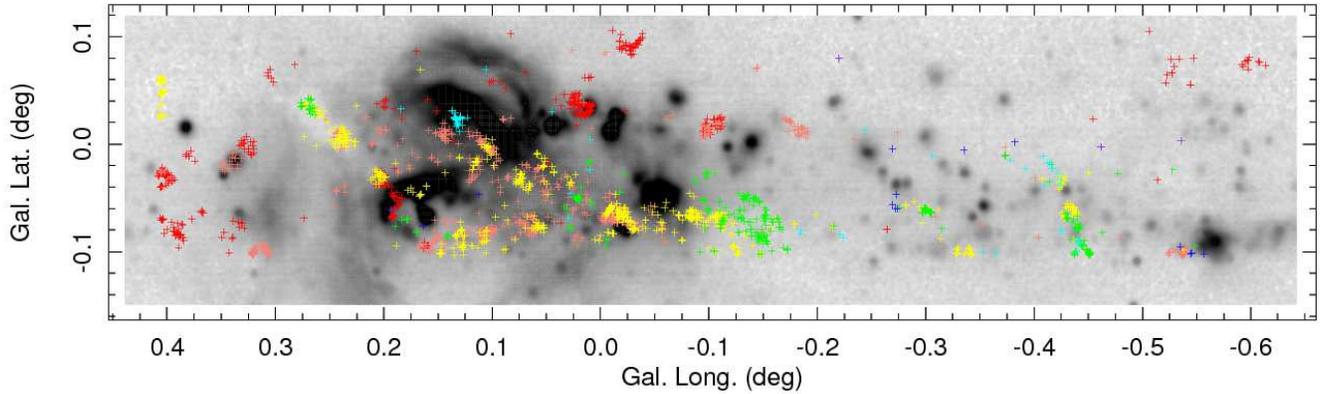}
\caption{Reverse grayscale of the 21 micron MSX image of region
surveyed with methanol masers shown as ``+''. 
Velocity is color coded, red$>$80 km $s^{-1}$, salmon=60 km $s^{-1}$,
yellow=30 km $s^{-1}$, green=0 km $s^{-1}$, cyan=-30 km $s^{-1}$, 
blue=-60 km $s^{-1}$, violet$<$-80 km $s^{-1}$, 
}
\label{maserMIRfig}
\end{figure*}

Figure \ref{maserMIRfig} shows the distribution of methanol masers on
the 21 micron Midcourse Space Experiment (MSX) image \cite{MSX}.
While some masers are in regions of star formation, many are 
{ coincident with }
infrared dark clouds (IRDC) and many have no apparent association with
either star forming regions or dark clouds.
{
While these masers have a widespread distribution, Figure
\ref{maserMIRfig} shows them to be highly clustered with clusters having
a relatively narrow range of velocities indicating that the cluster
arises from a single cloud or group of clouds.
}

\begin{figure*}
\includegraphics[width=10.0in,angle=0]{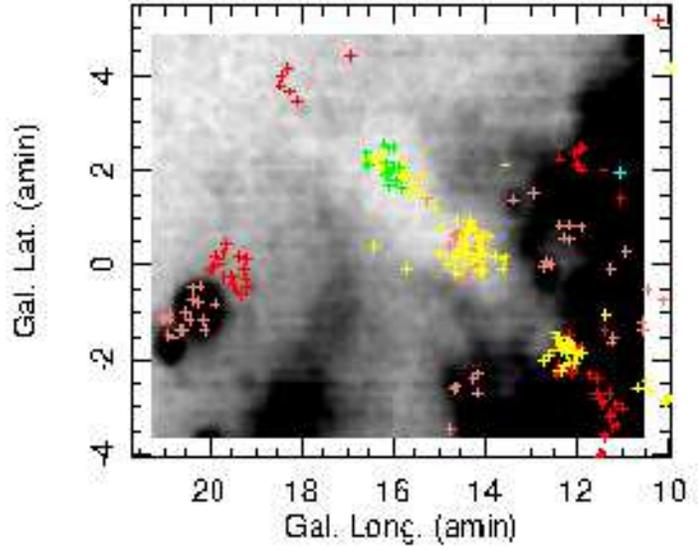}
\caption{Closeup of the reverse grayscale of the 21 micron MSX image
of an IRDC (``The Brick'') with methanol masers shown as crosses.
Velocity is color coded, red$>$80 km $s^{-1}$, salmon=60 km $s^{-1}$,
yellow=30 km $s^{-1}$, green=0 km $s^{-1}$, cyan=-30 km $s^{-1}$.
}
\label{IRDCfig}
\end{figure*}

A detailed view of the region of an IRDC, ``The Brick'', is given in Figure
\ref{IRDCfig}.
This figure shows two clusters of methanol masers in prominent IR dark
regions with radial velocities around 40 and 0 km s$^{-1}$.
On the left side of this figure is the star forming region 
IRAS1743-2838 (0.33-0.02) \citep{avedisova} with a cluster of masers near 76
km s$^{-1}$.

 We note a large number of new masers compared to the number of 
36.2 GHz maser candidates found in our earlier continuum observations
\citep{Paper1}.  
{ 
This is expected as the current data have the same sensitivity as
the earlier data but 1/16 of the channel width.  The narrow maser
features would be reduced by a factor of $\approx$16 in the older data
due to the wider "continuum" channels.
}
{ Mills et al. (2015) observe the Brick cloud (G0.253+0.016) and also
find an order of magnitude more methanol masers than the number of
candidate masers identified by Yusef-Zadeh et al. (2013) as well as
differences in source lists.
There are differences in the spatial resolution of these
studies (1.6''$\times$1.5'' vs. 1.8''$\times$0.7'') and in the methods
in which  masers were identified in Mills et al. (2015) and
Yusef-Zadeh et al. (2013).
Mills et al. (2015) apply the Clumpfind algorithm to identify maser
sources and compare the positions of their cataloged sources with our
earlier measurements and find discrepancies.
In addition, the discrepancies in the identification of some masers
toward this cloud in Yusef-Zadeh et al. (2013) and Mills et al. (2015)
could be  due to different sensitivity (30 sec. vs. 24 min on source
integration) and frequency resolution (16.6 km s$^{-1}$ vs. $\sim$ 1
km s$^{-1}$).  
Variations in maser brightness cannot be ruled out.
Yusef-Zadeh et al. (2013) reported maser candidates based on continuum
observation and the maser sources needed to be confirmed with higher
resolution spectral line observations.  
Present high resolution spectroscopic observations show maser
sources.}  
A number of sources identified in Mills et al. (2015) and 
Yusef-Zadeh et al. (2013) have no counterparts in our present {
high resolution} survey. 
These studies indicate that the spatial resolution is critical for
separating masers from quasi-thermal sources
{ which are resolved out in the high resolution A configuraton observations.}



Another Galactic center cloud that  shows 36.2 GHz methanol masers is 
the Sgr A East molecular cloud.
Sjouwerman et al. (2010) identify 10 methanol masers at the
interaction site between the Sgr A East supernova remnant, the 50 km
s$^{-1}$   molecular cloud and  the 20 km $^{-1}$ molecular cloud. 
We compared the methanol  masers from Sjouwerman et al. (2010) 
with those listed  in Table  3 and found that they  agree well
in both velocity and position.  




\subsection{1.7 GHz Continuum Results}
The continuum images are strongly disturbed by the artifacts of the
strong emission { in the neighborhood of the Galactic center (HII
regions, nonthermal arcs, filaments, etc.)} which is nearly completely resolved by
the high resolution of the data.
However, it is possible to identify a number of small and isolated
sources and these are given in Table \ref{ContLCat}.
Due to the strongly variable nature of the images, it is not possible
to characterize the selection criteria such as the minimum flux density.

\subsection{OH (1612, 1665/7, 1720 MHz) Masers}
\begin{figure*}
\centerline{
\includegraphics[width=2.5in,angle=-90]{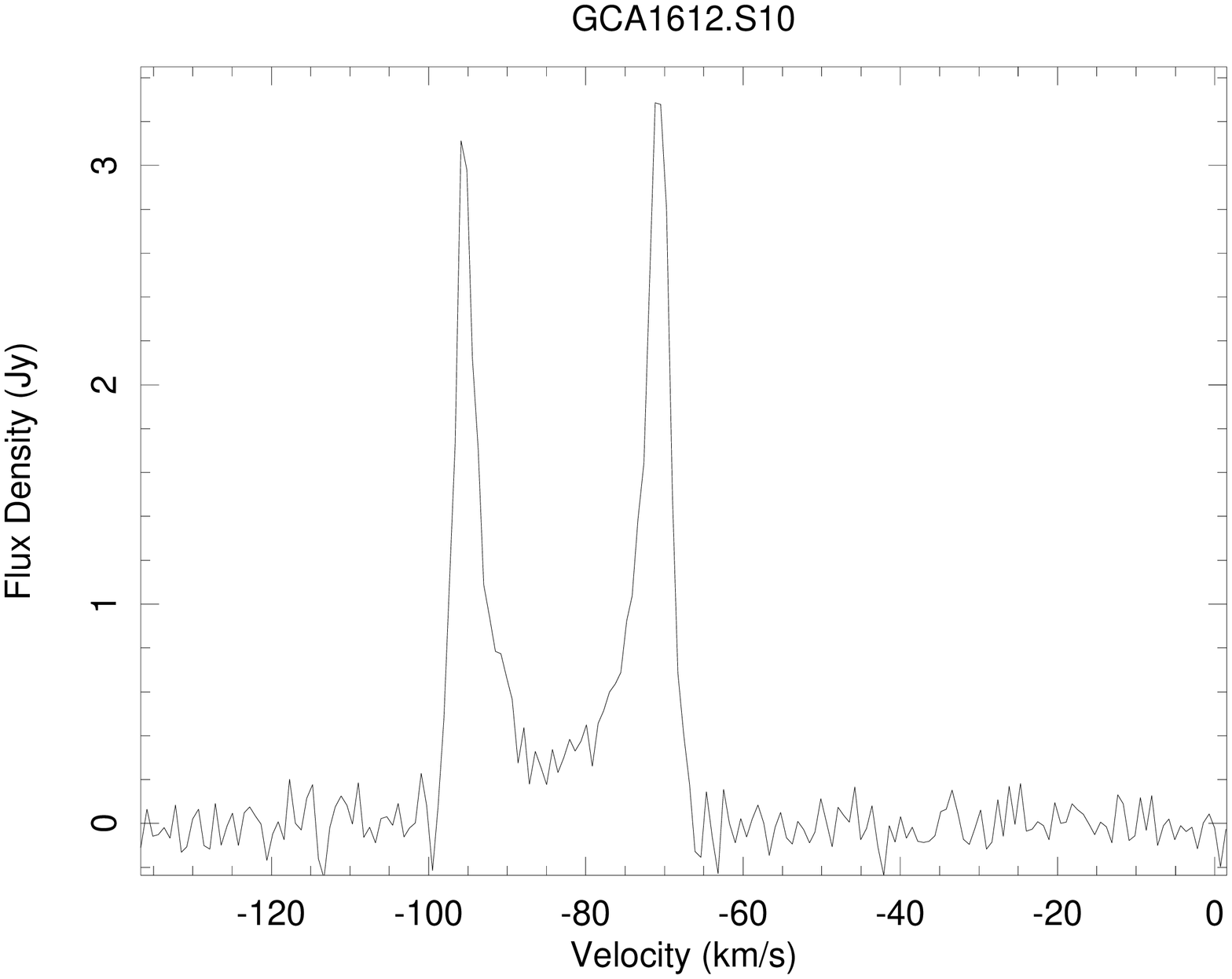}
\includegraphics[width=2.5in,angle=-90]{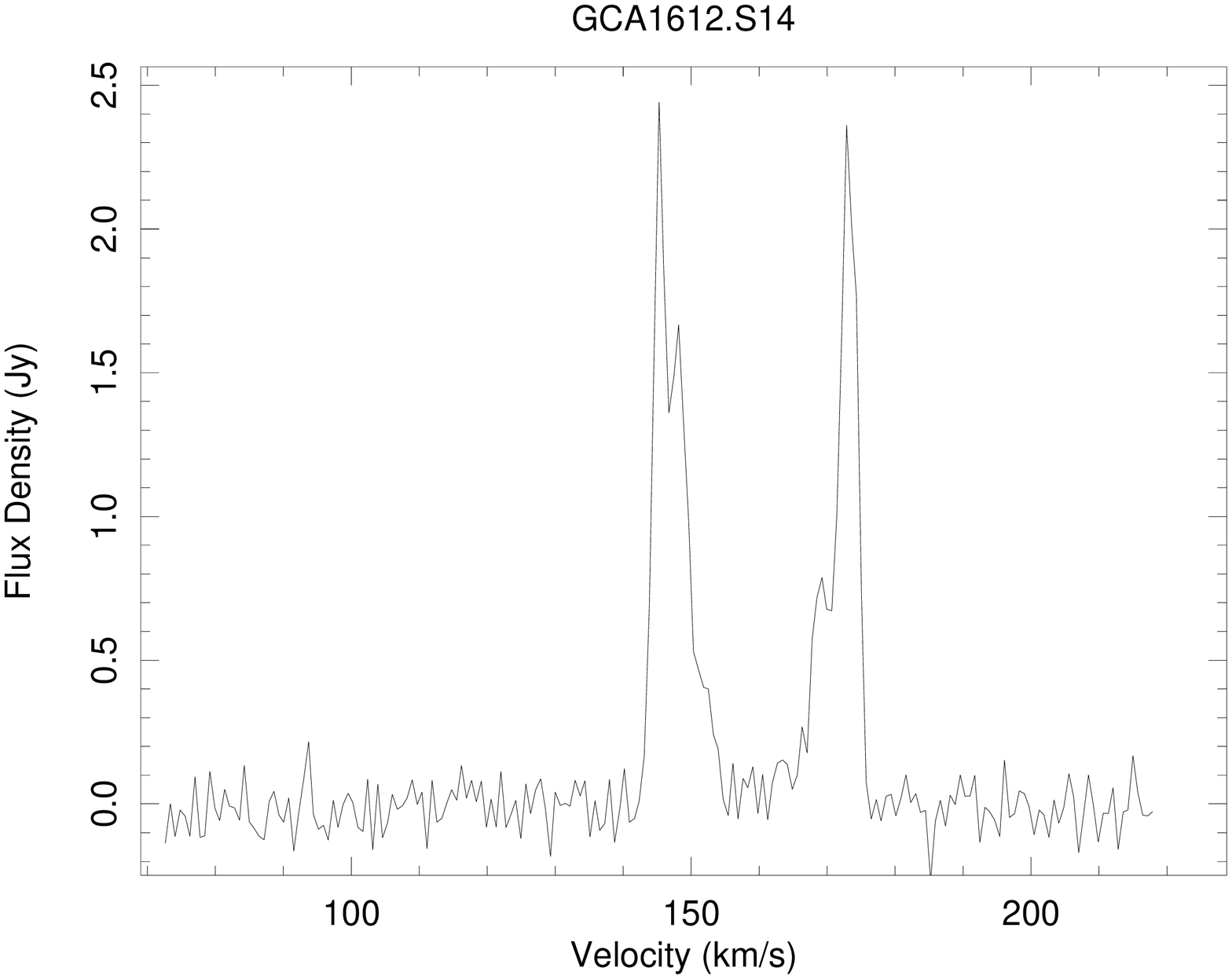}
}
\centerline{
\includegraphics[width=2.5in,angle=-90]{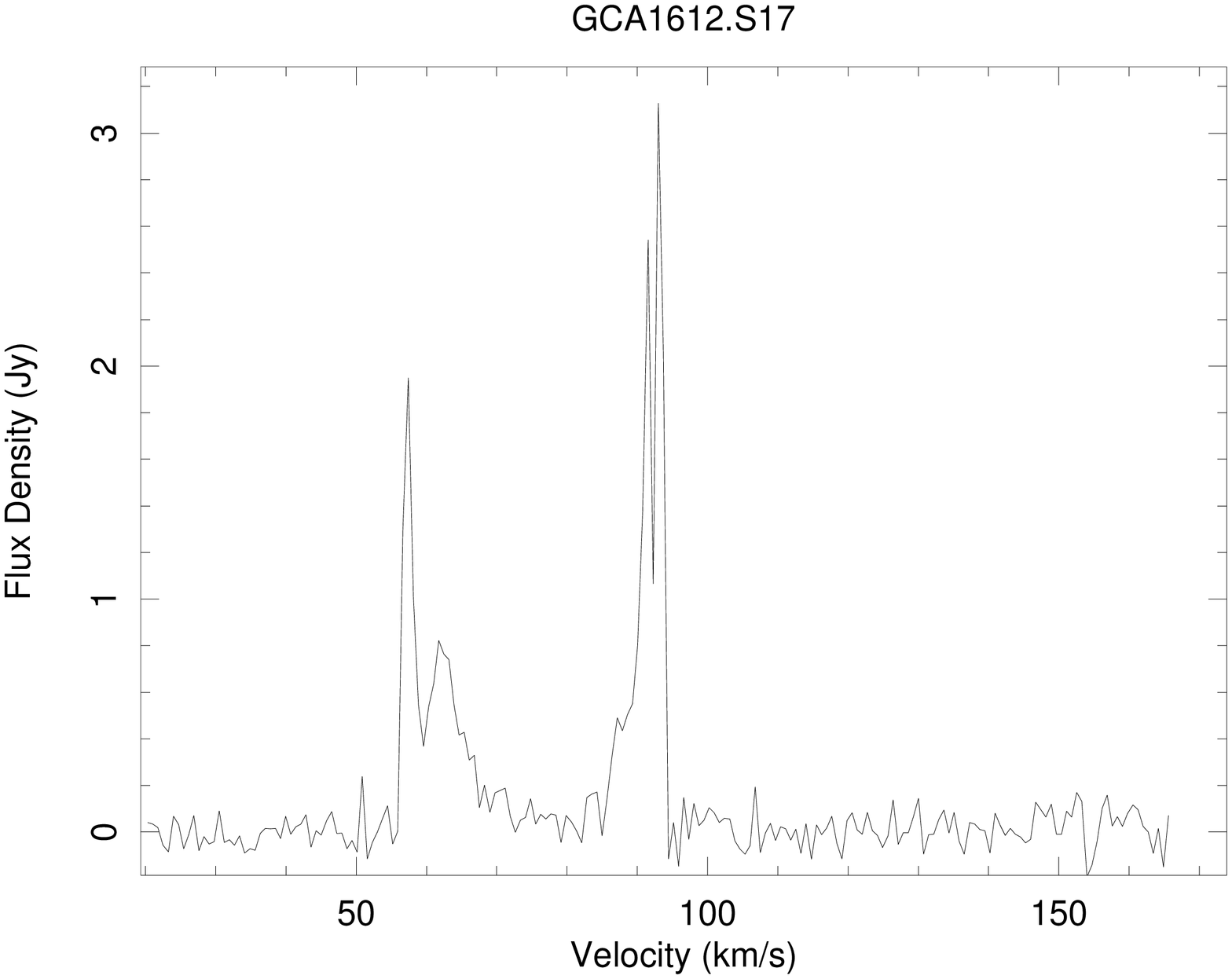}
\includegraphics[width=2.5in,angle=-90]{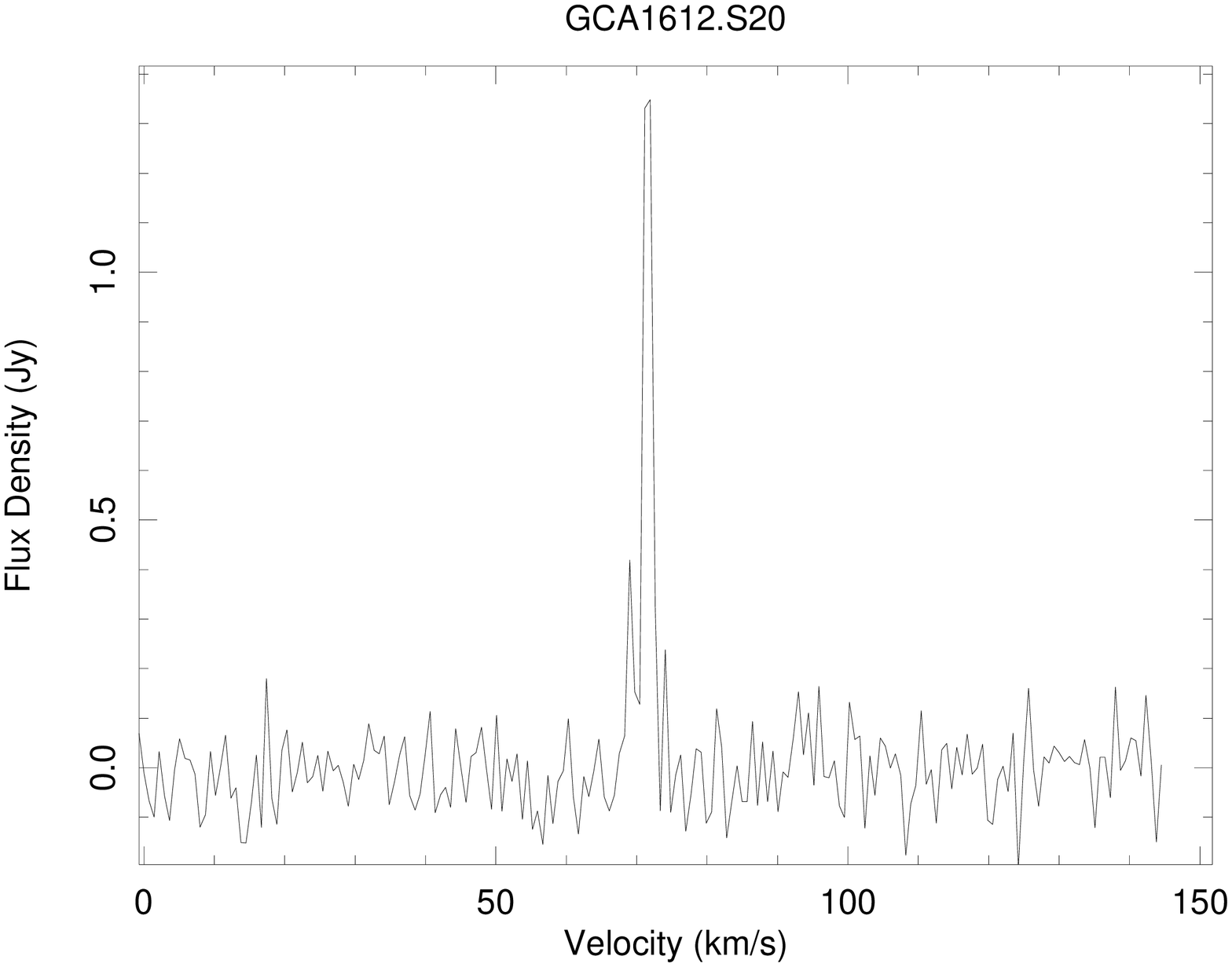}
}
\centerline{
\includegraphics[width=2.5in,angle=-90]{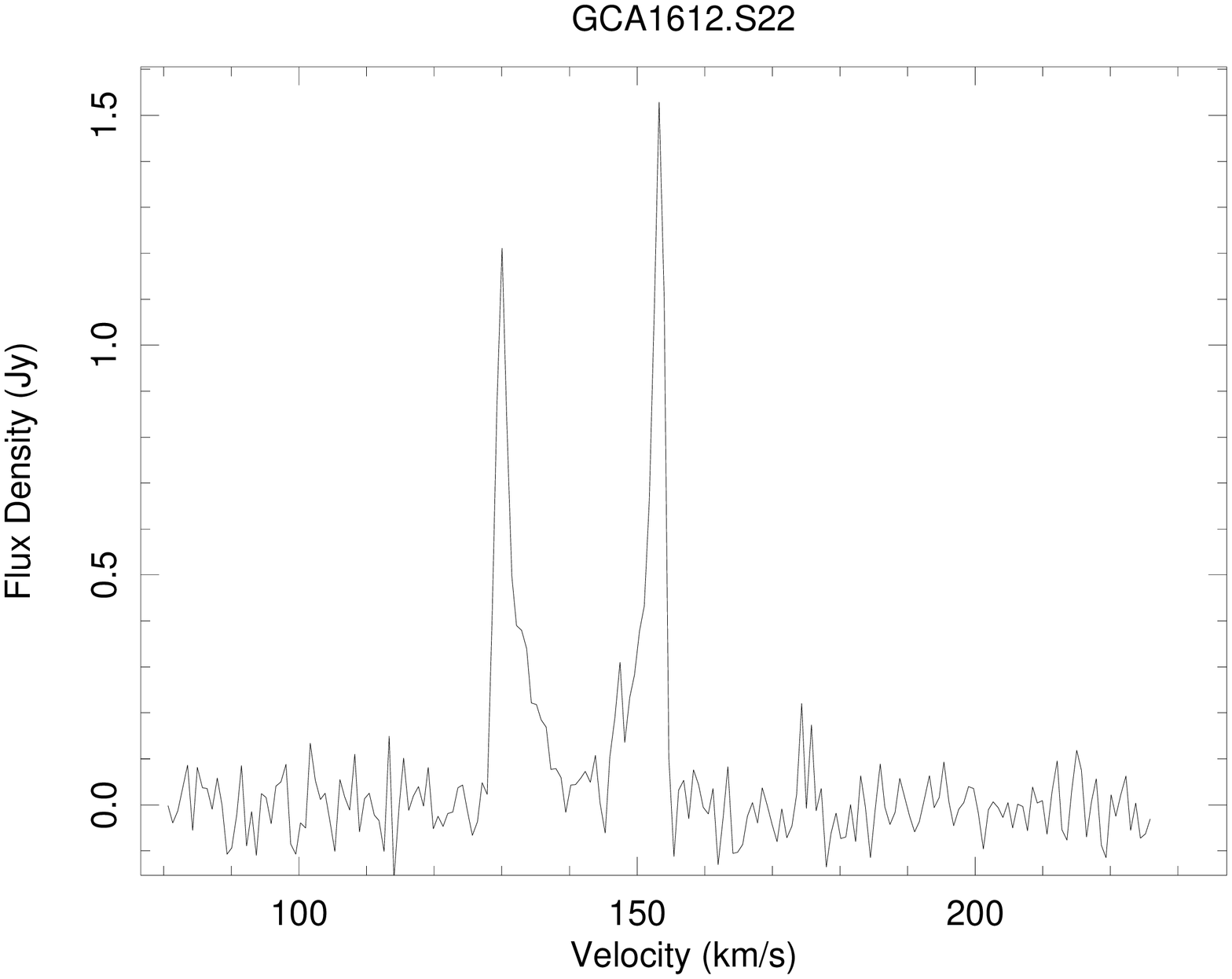}
\includegraphics[width=2.5in,angle=-90]{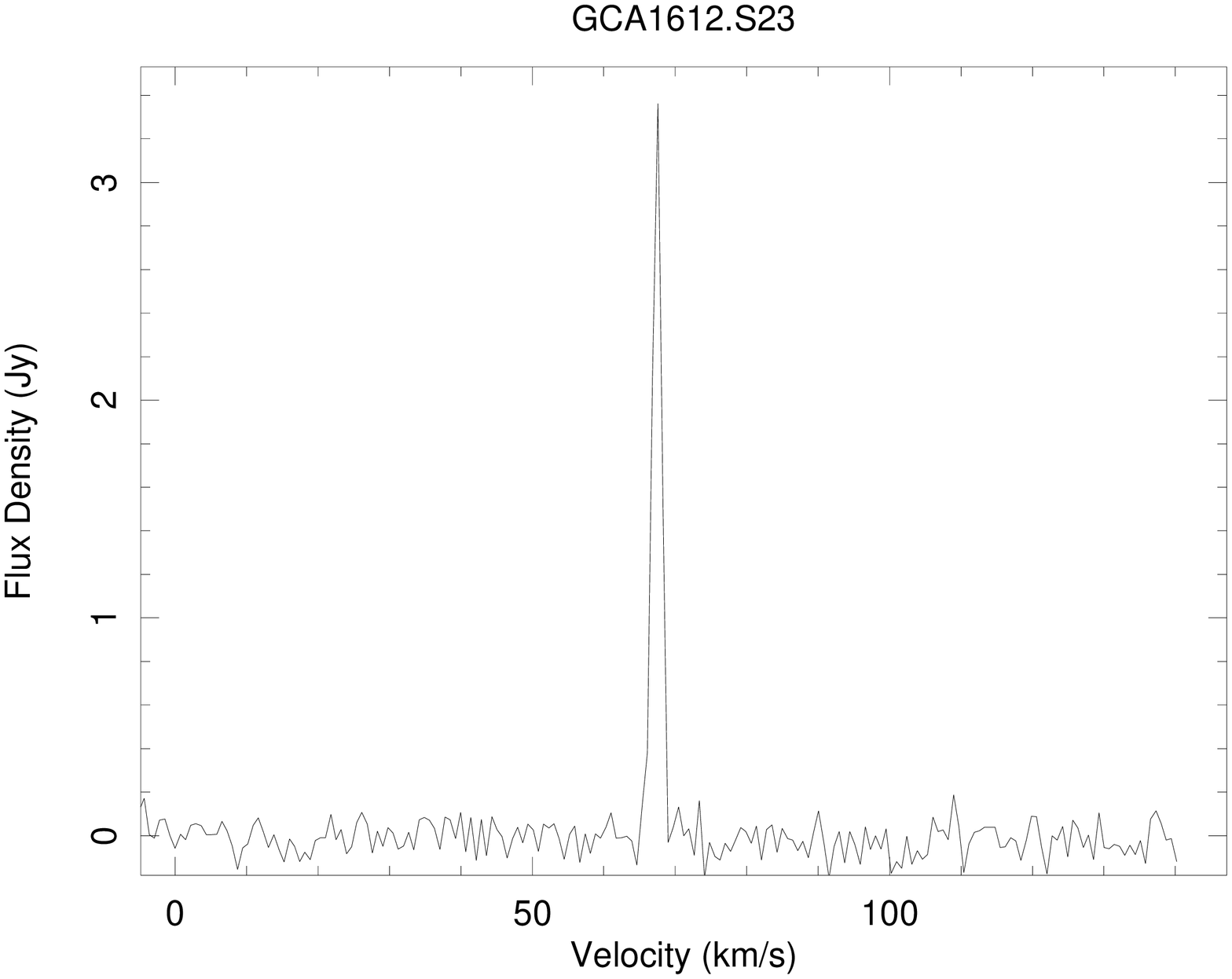}
}
\caption {Sample 1612 MHz OH maser spectra.
The source { name} is given above each plot.}
\label{OH1612Spectfig}
\end{figure*}

Masers were identified in the combined spectral cubes in the manner
described in Section \ref{Finding} with a minimum peak flux density
the greater of 8 times the local RMS and 15\% of the channel peak flux
density. 
Maser velocity and width were then determined from a moment analysis.
A list of the 1612 MHz OH sources is given in Table
\ref{OH1612MaserCatalog} with 13 columns indicating  
the source { name}, celestial coordinates, Galactic coordinates, the
flux density, the peak velocity, the line-width, spectrum type and the
lower limit on the brightness temperature.   
The column ``Type'' identifies double (``D'')  and single (``S'')
peaked spectra.  
Sample spectra are shown in Figure \ref{OH1612Spectfig}.
Spectra are in two distinct types, double peaked (``D'') typical of
evolved stars and single peaked (``S'') with the former dominating the
sample.  
Velocities and widths given in Table \ref{OH1612MaserCatalog} for type
``D'' sources are those of the stronger component; the typical total
velocity width is $\sim$40 km s$^{-1}$.

\begin{figure*}
\centerline{
\includegraphics[width=2.5in,angle=-90]{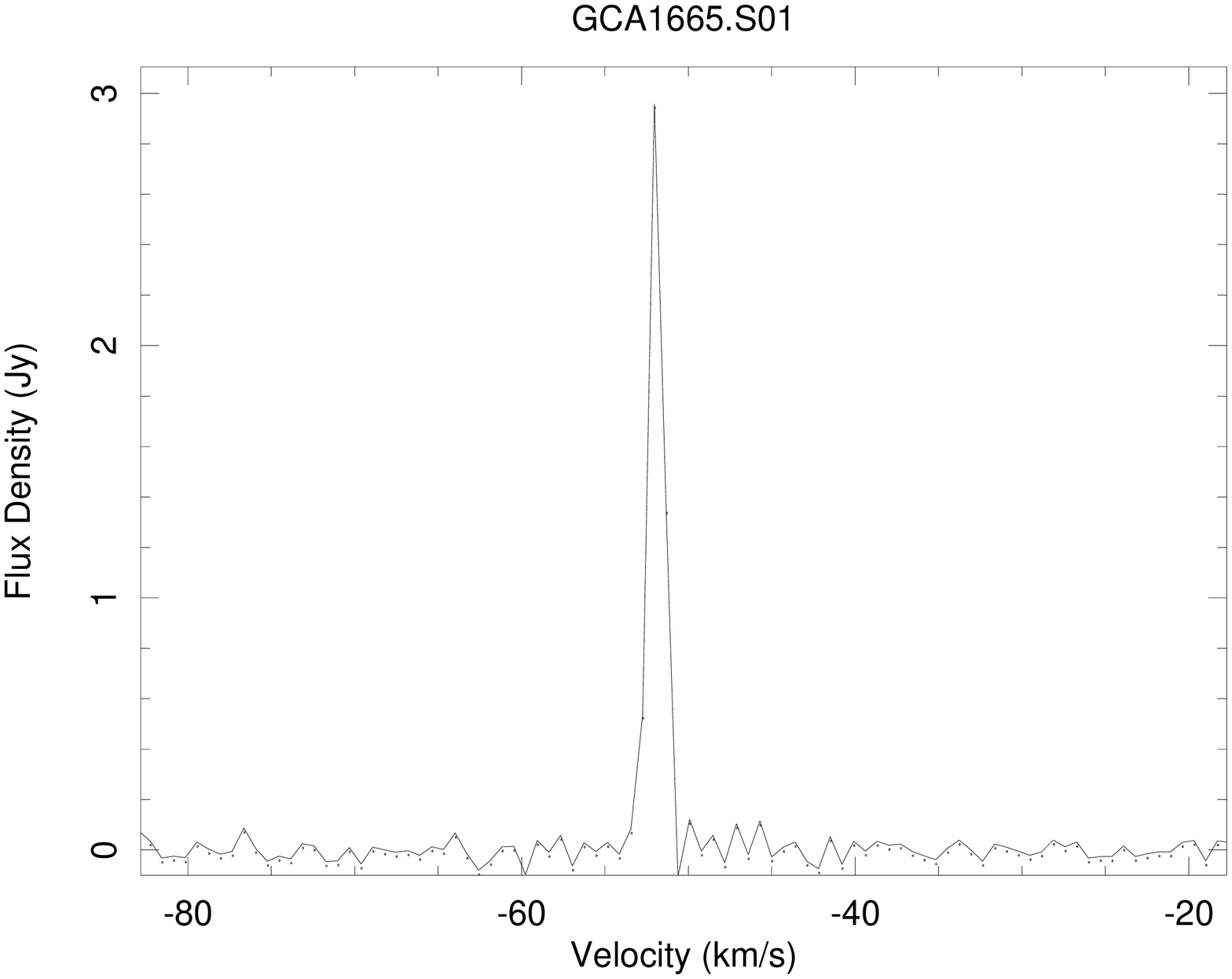}
\includegraphics[width=2.5in,angle=-90]{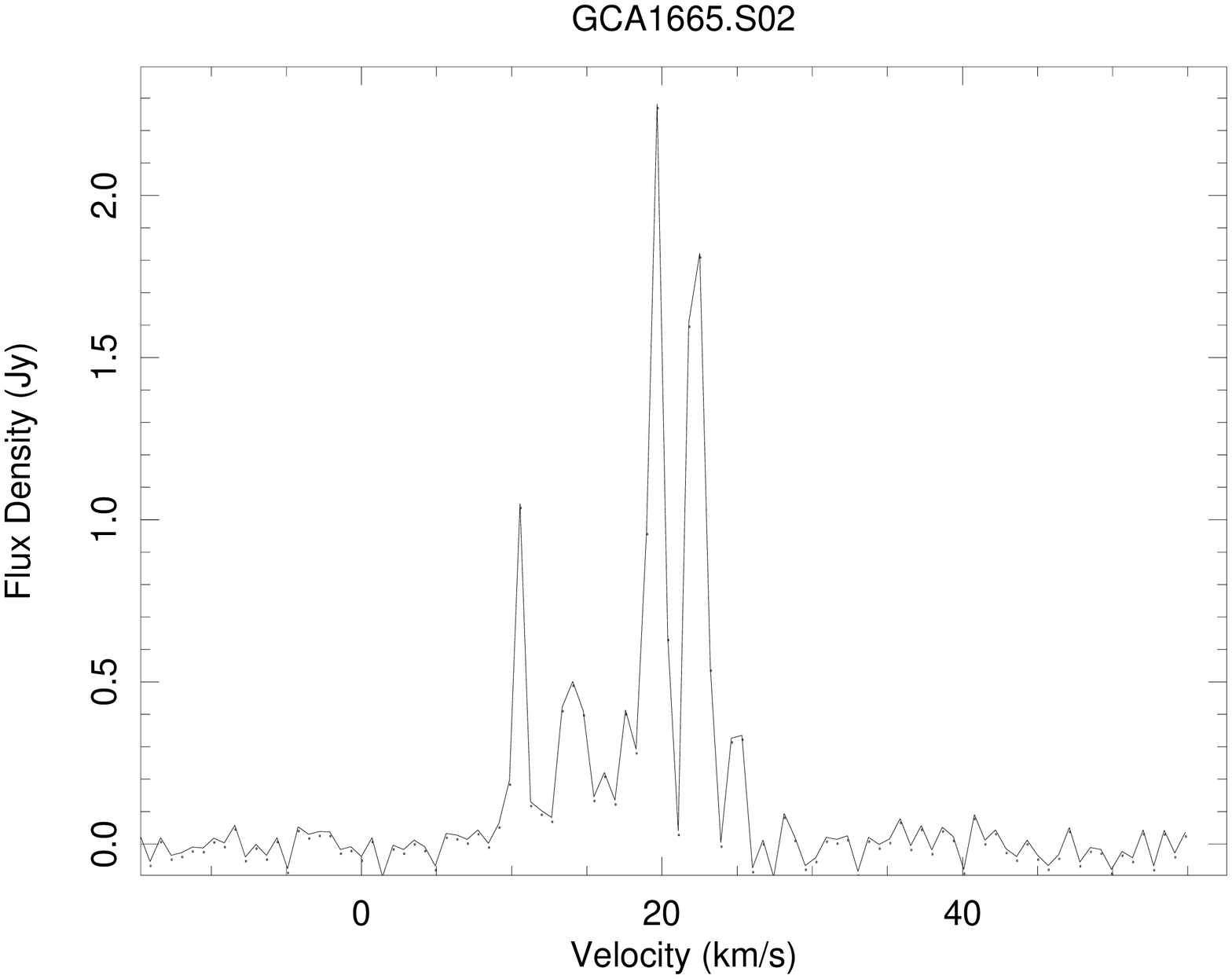}
}
\centerline{
\includegraphics[width=2.5in,angle=-90]{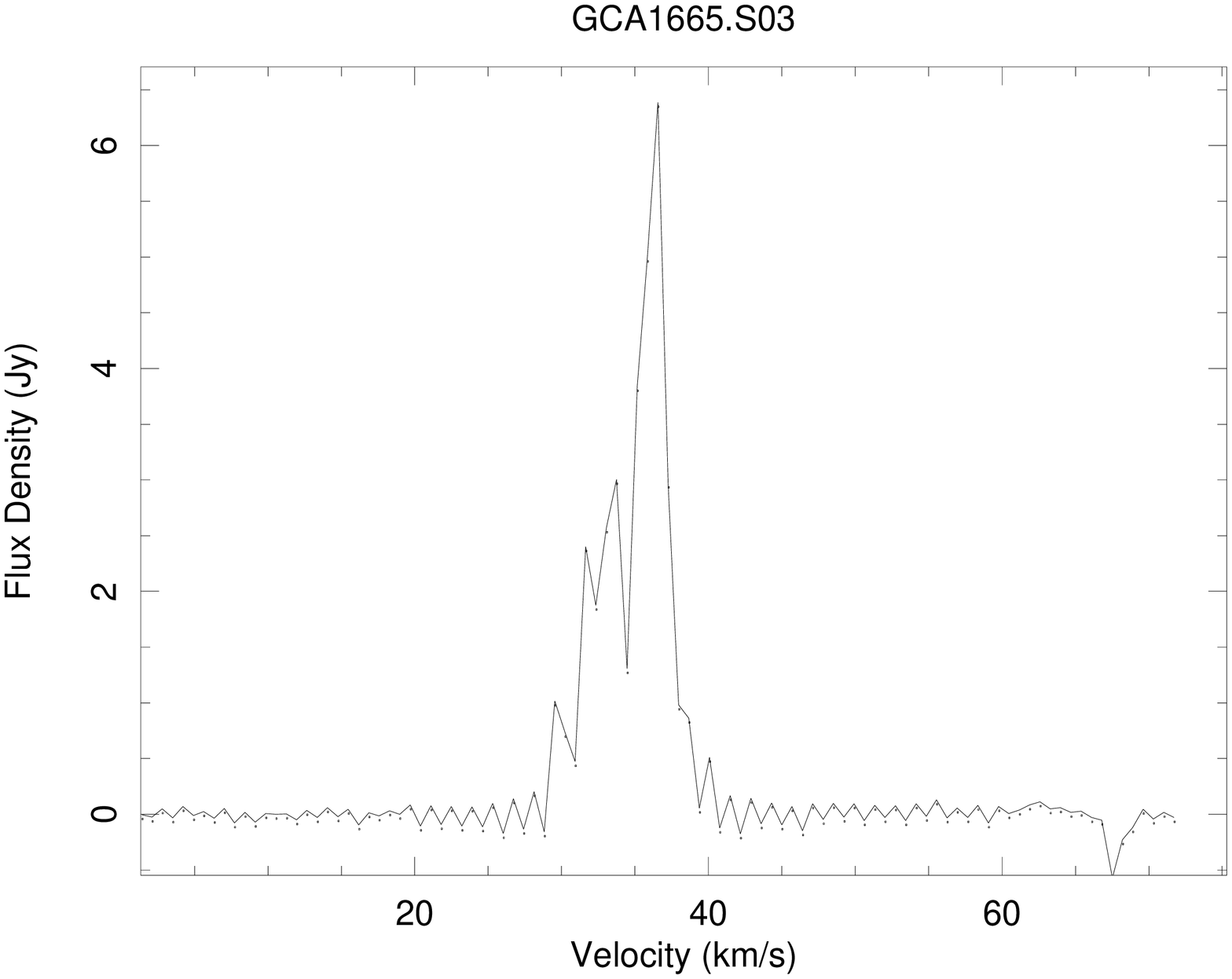}
\includegraphics[width=2.5in,angle=-90]{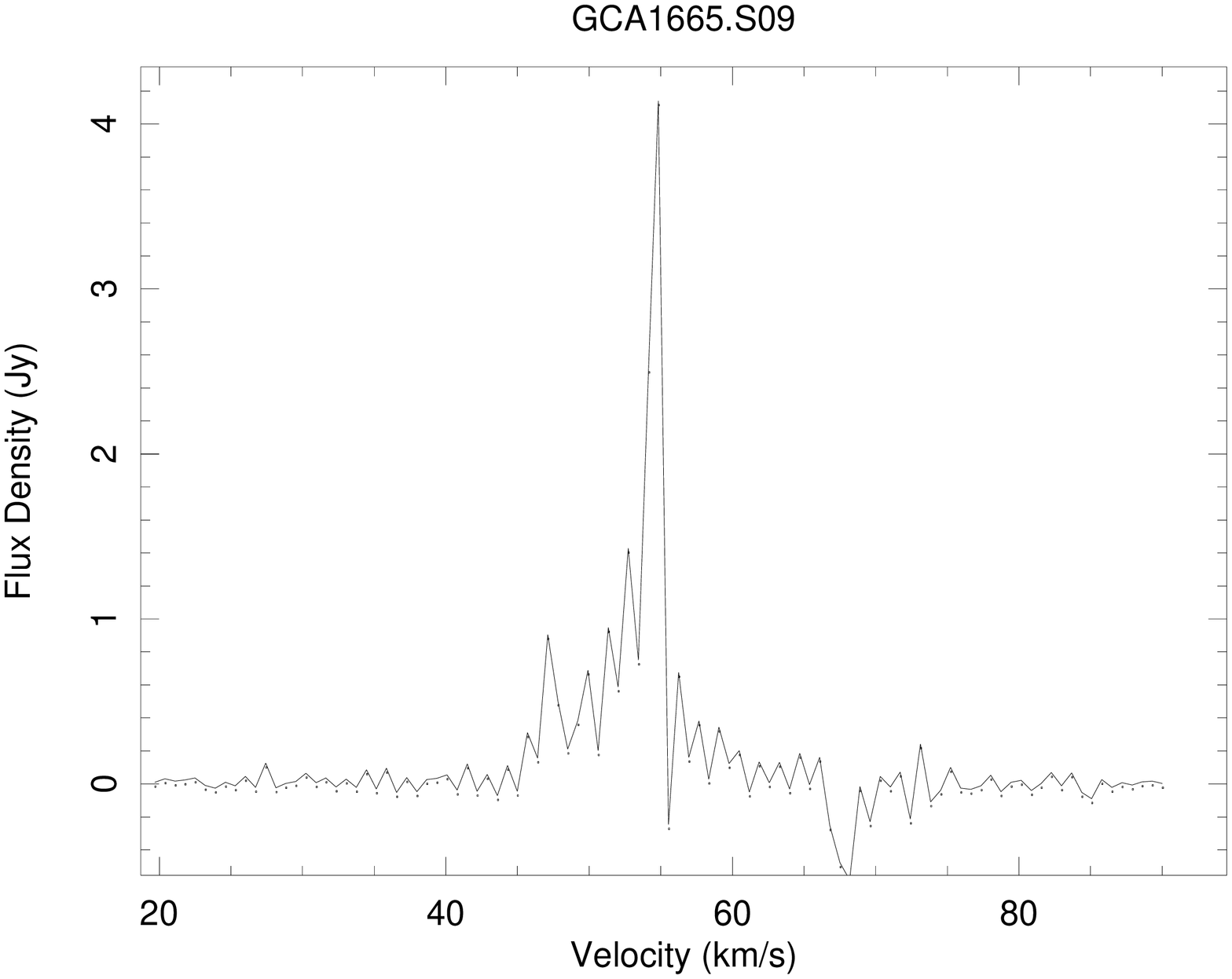}
}
\centerline{
\includegraphics[width=2.5in,angle=-90]{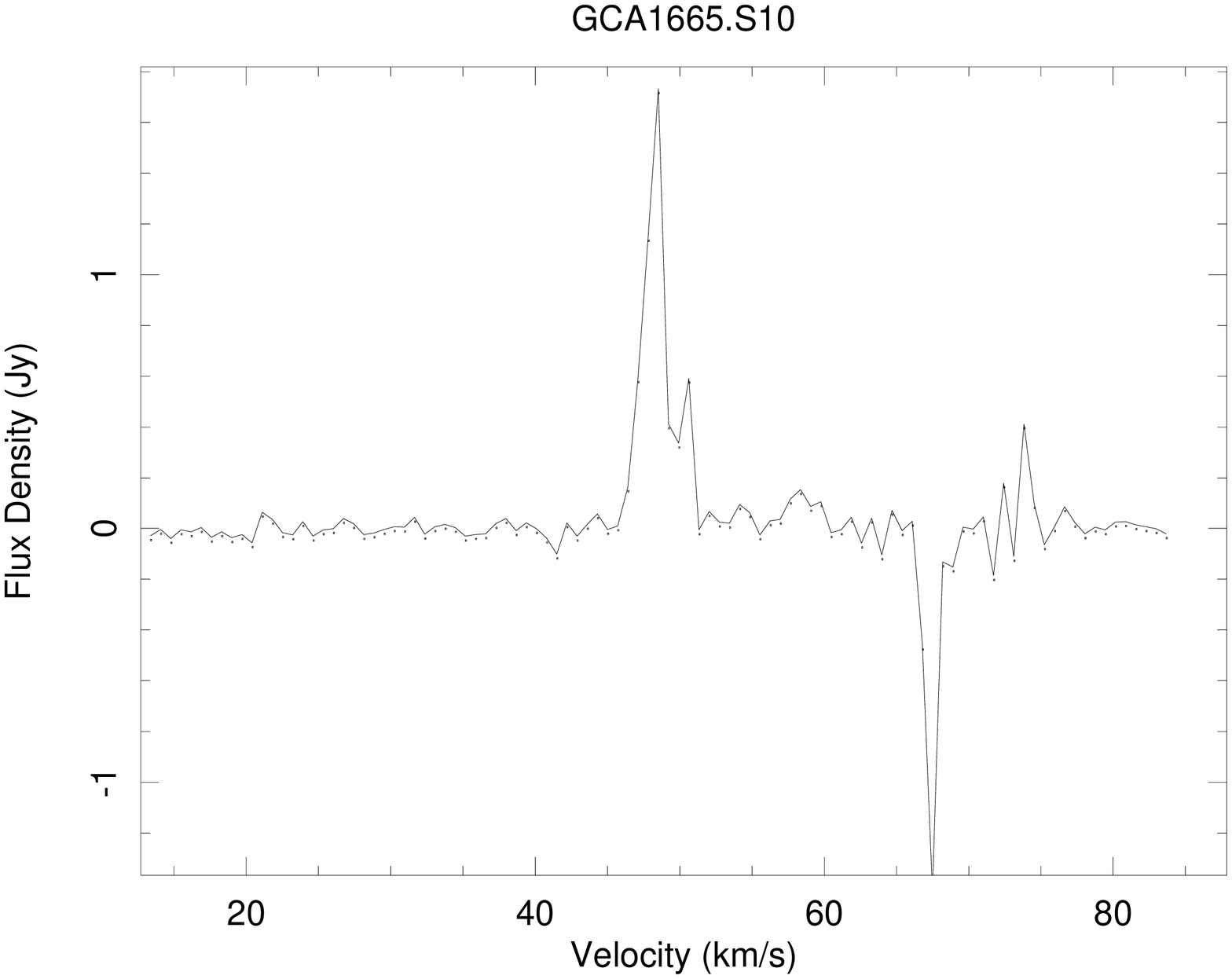}
\includegraphics[width=2.5in,angle=-90]{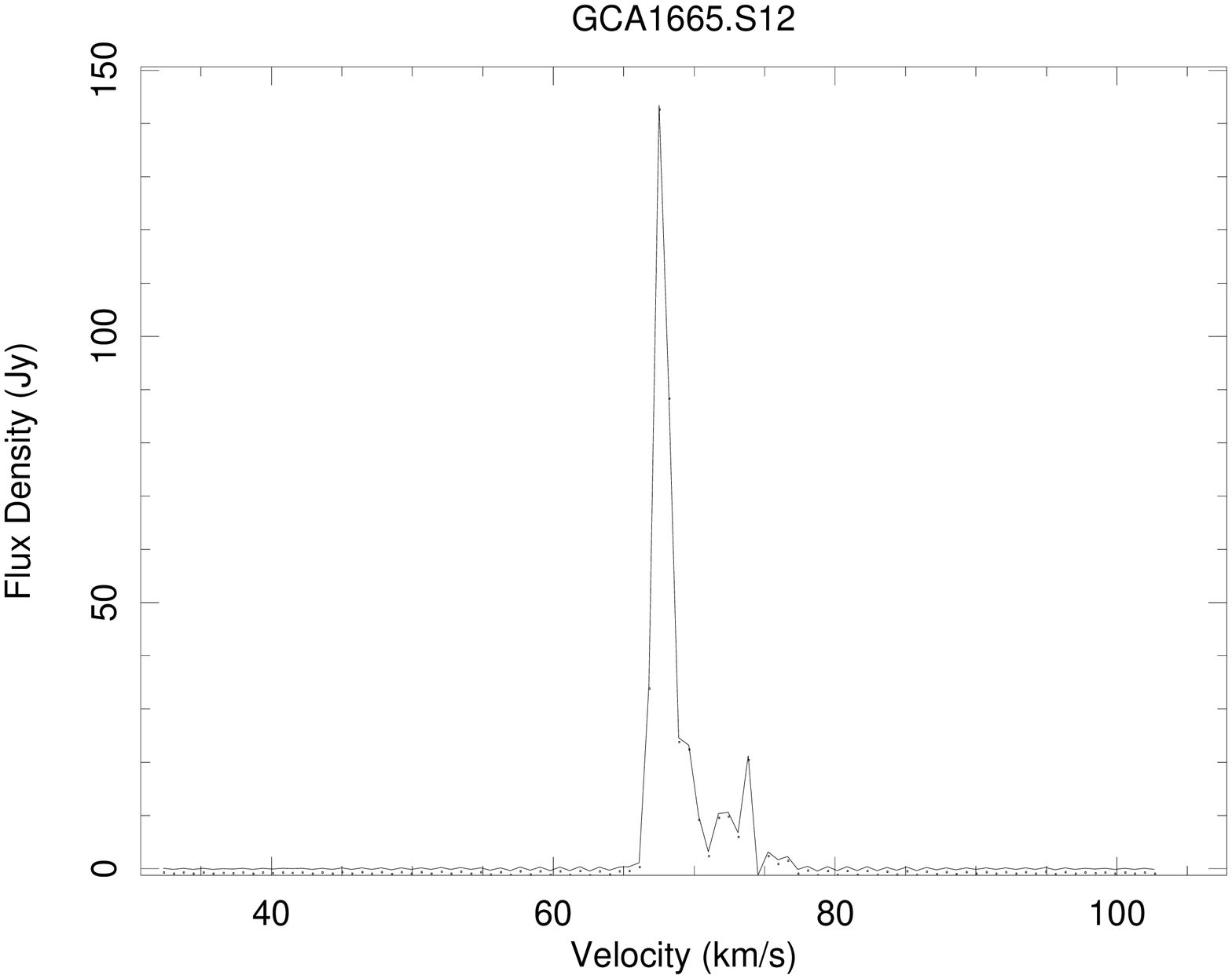}
}
\caption {Sample 1665 MHz OH maser spectra.}
\label{OH1665Spectfig}
\end{figure*}
\begin{figure*}
\centerline{
\includegraphics[width=2.5in,angle=-90]{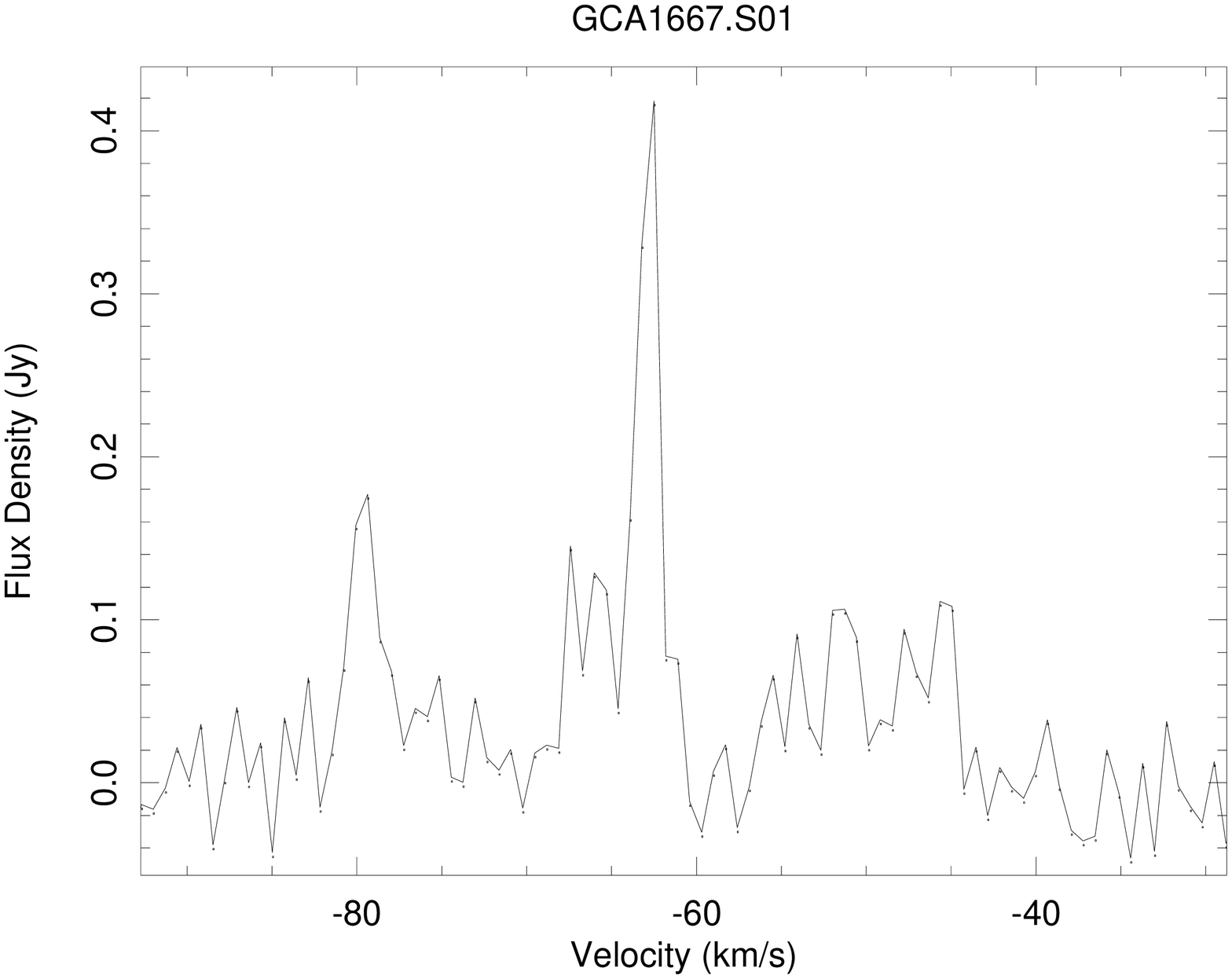}
\includegraphics[width=2.5in,angle=-90]{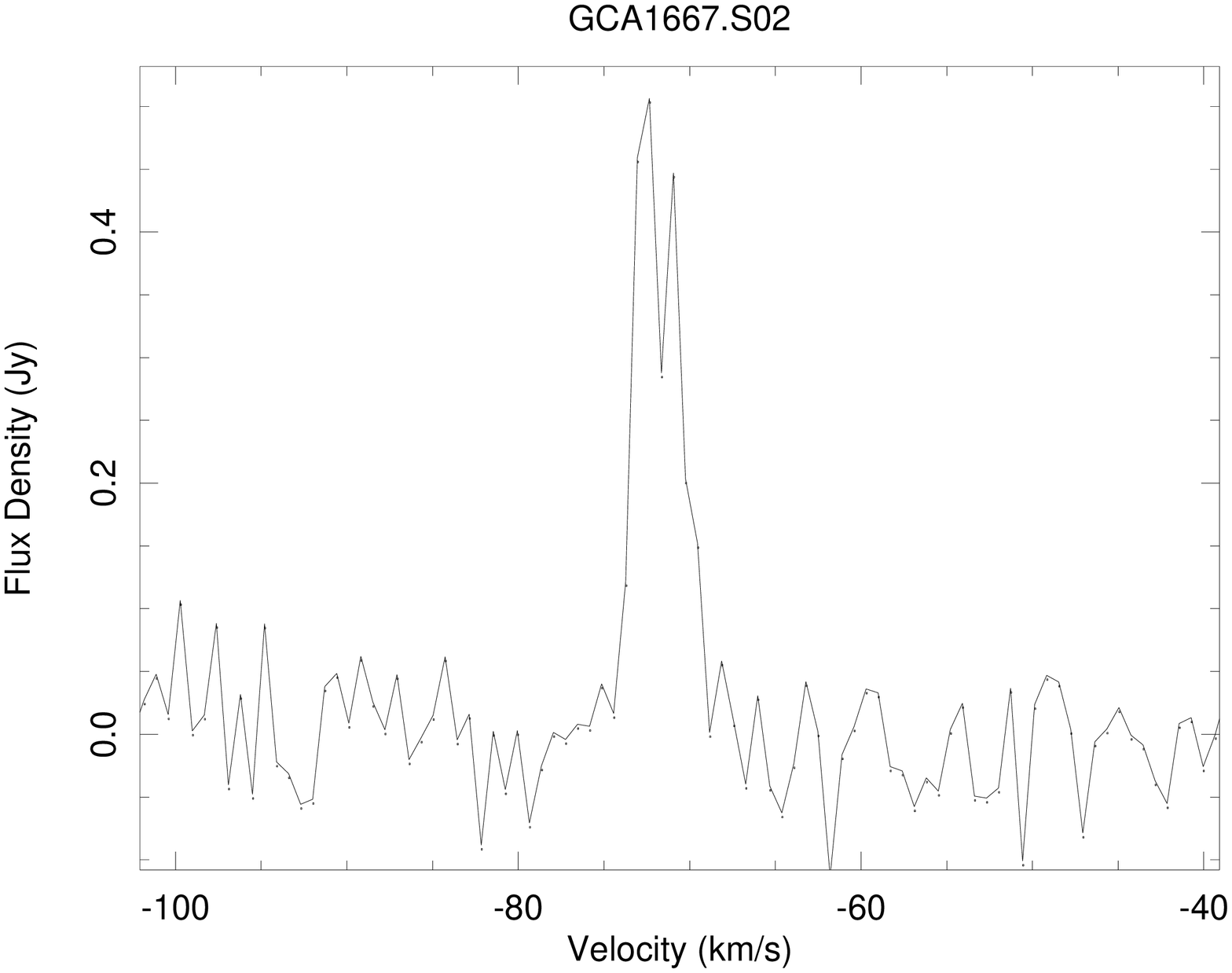}
}
\centerline{
\includegraphics[width=2.5in,angle=-90]{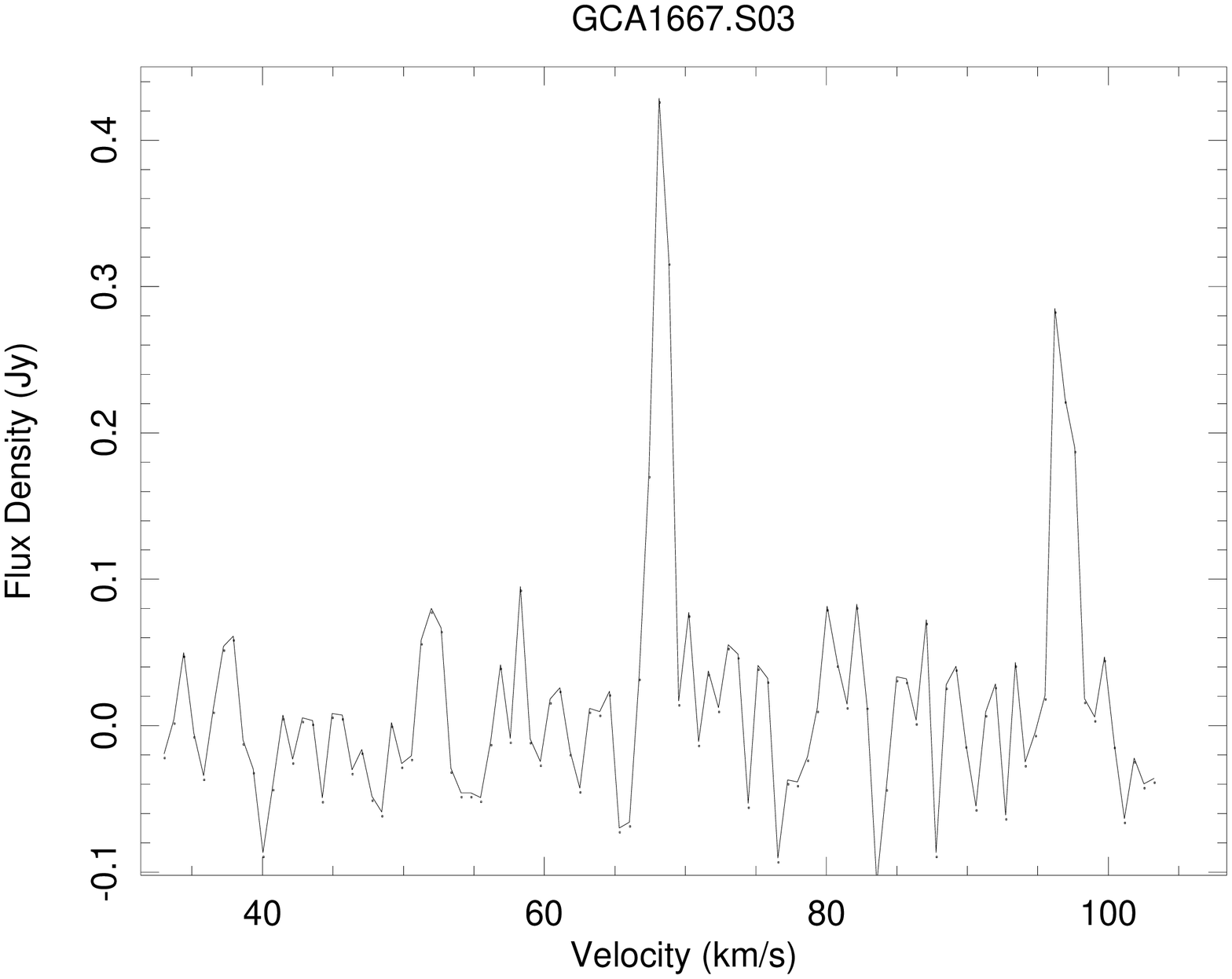}
\includegraphics[width=2.5in,angle=-90]{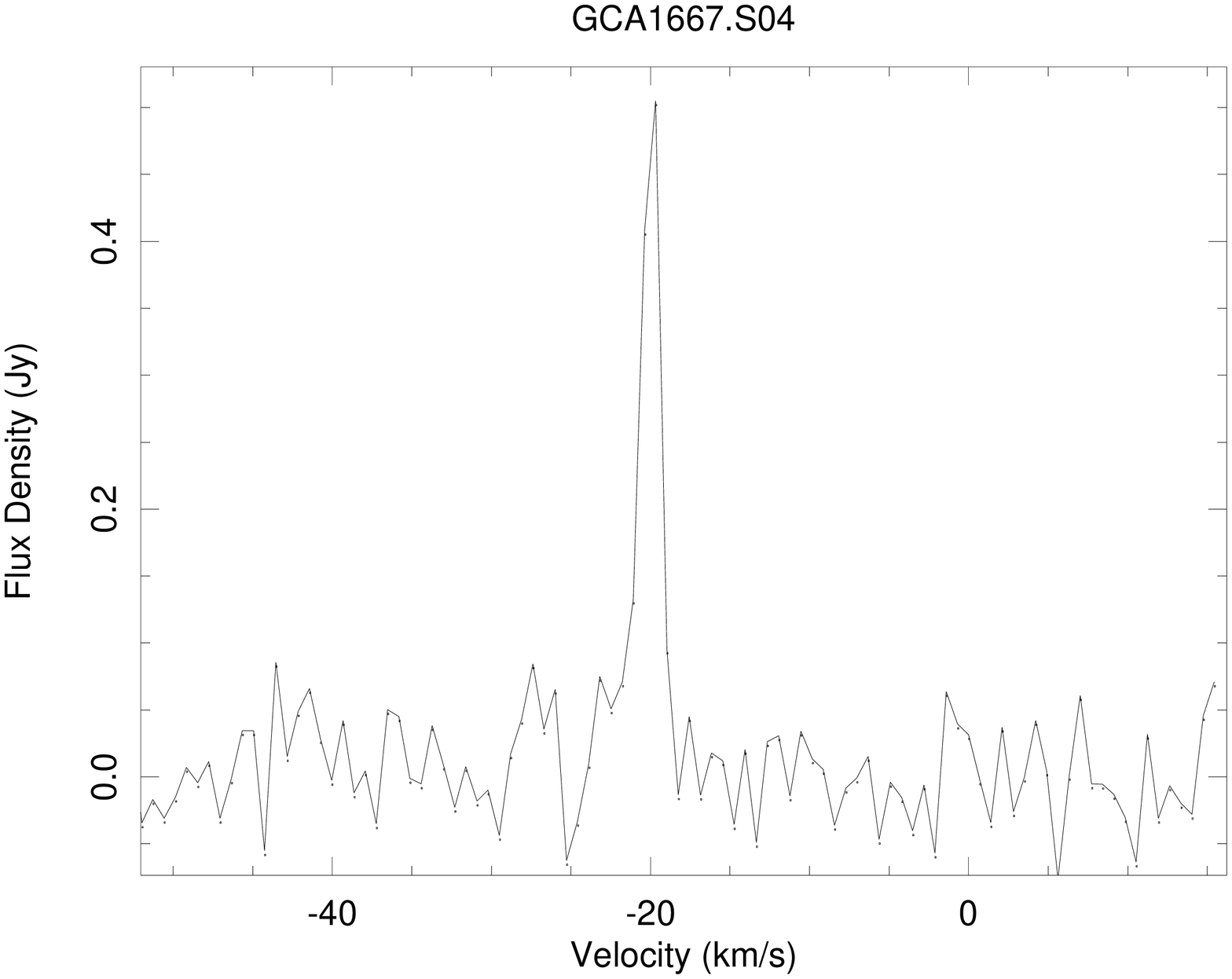}
}
\centerline{
\includegraphics[width=2.5in,angle=-90]{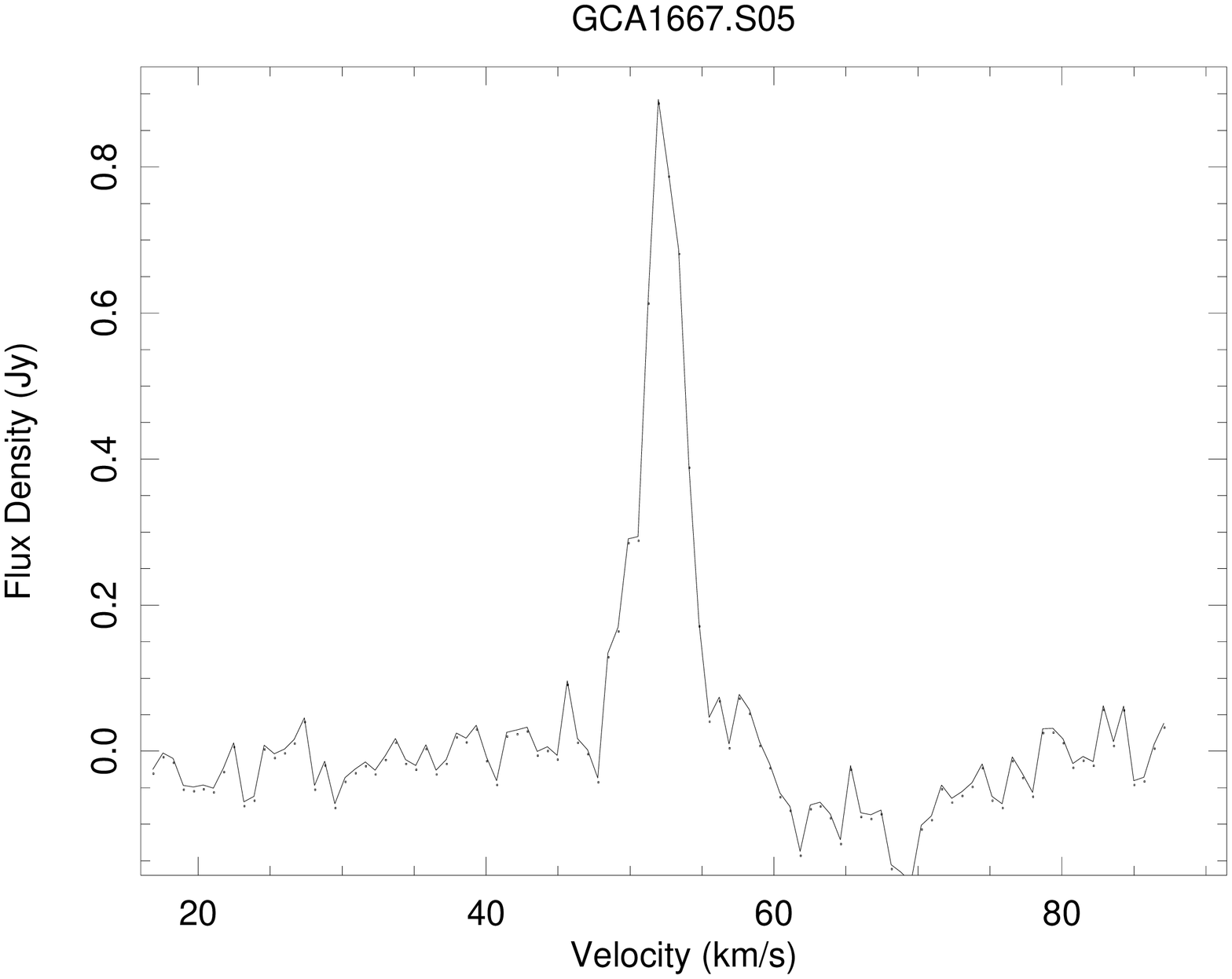}
}
\caption {1667 MHz OH maser spectra}
\label{OH1667Spectfig}

\end{figure*}
\begin{figure*}
\centerline{
\includegraphics[width=2.5in,angle=-90]{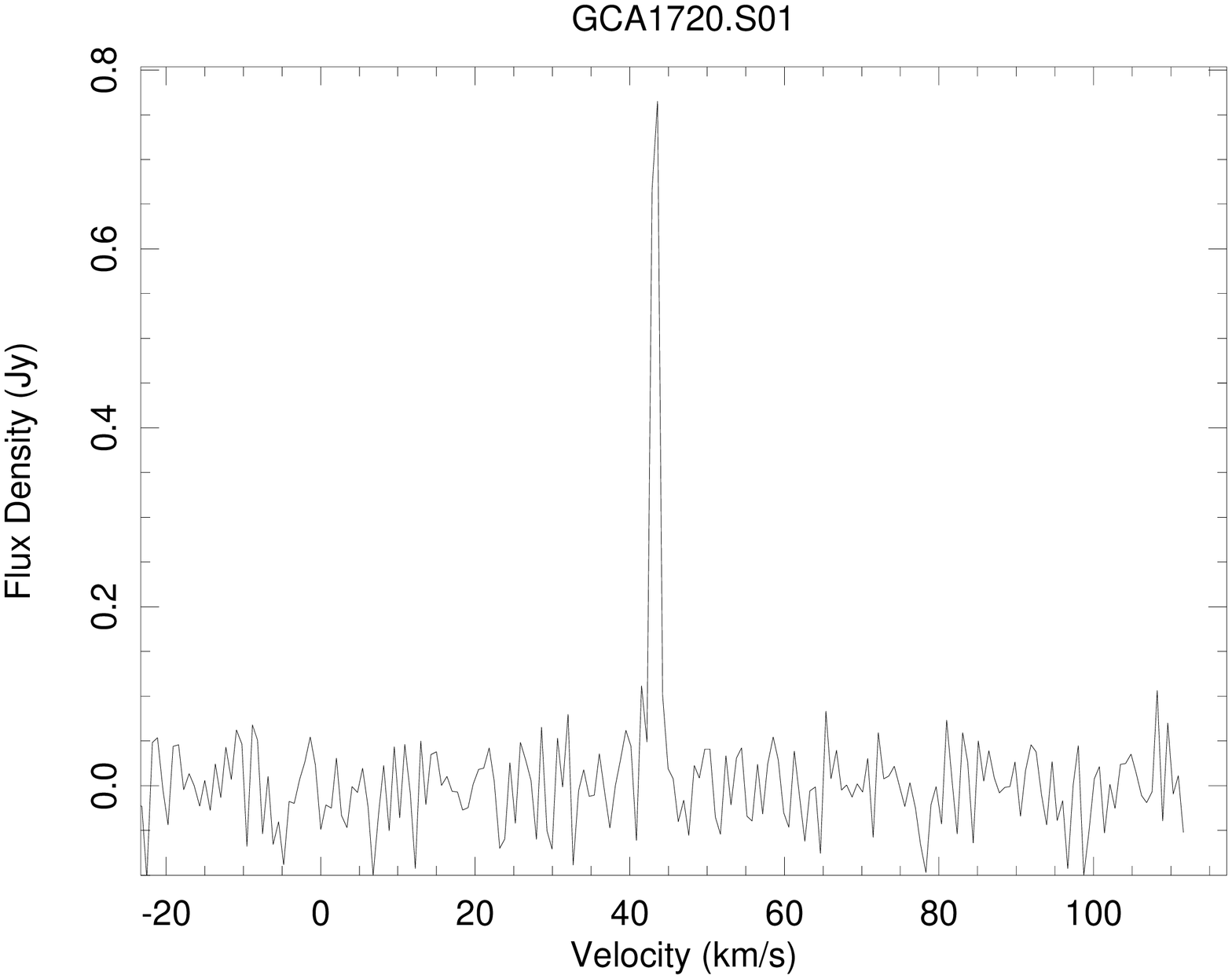}
\includegraphics[width=2.5in,angle=-90]{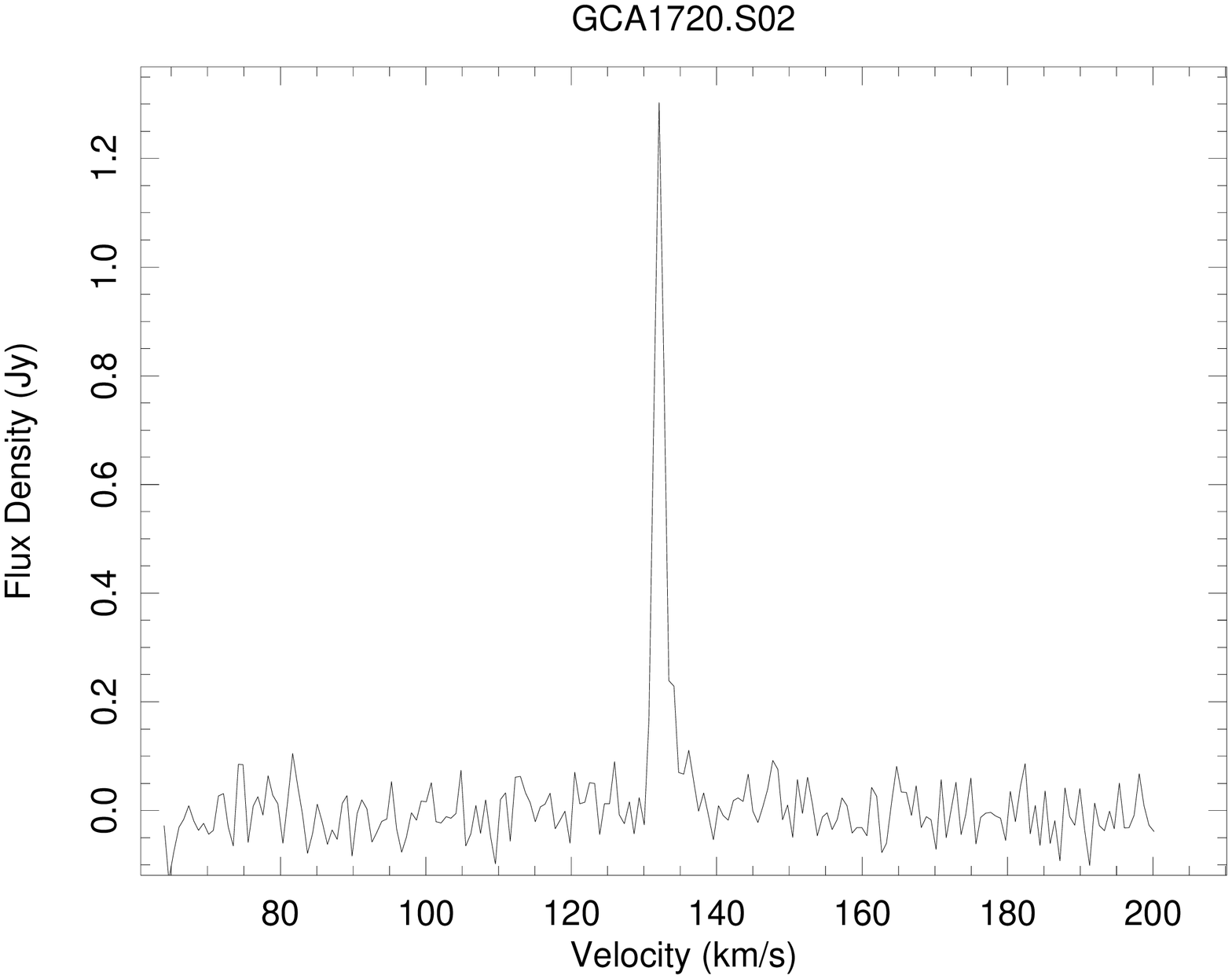}
}
\centerline{
\includegraphics[width=2.5in,angle=-90]{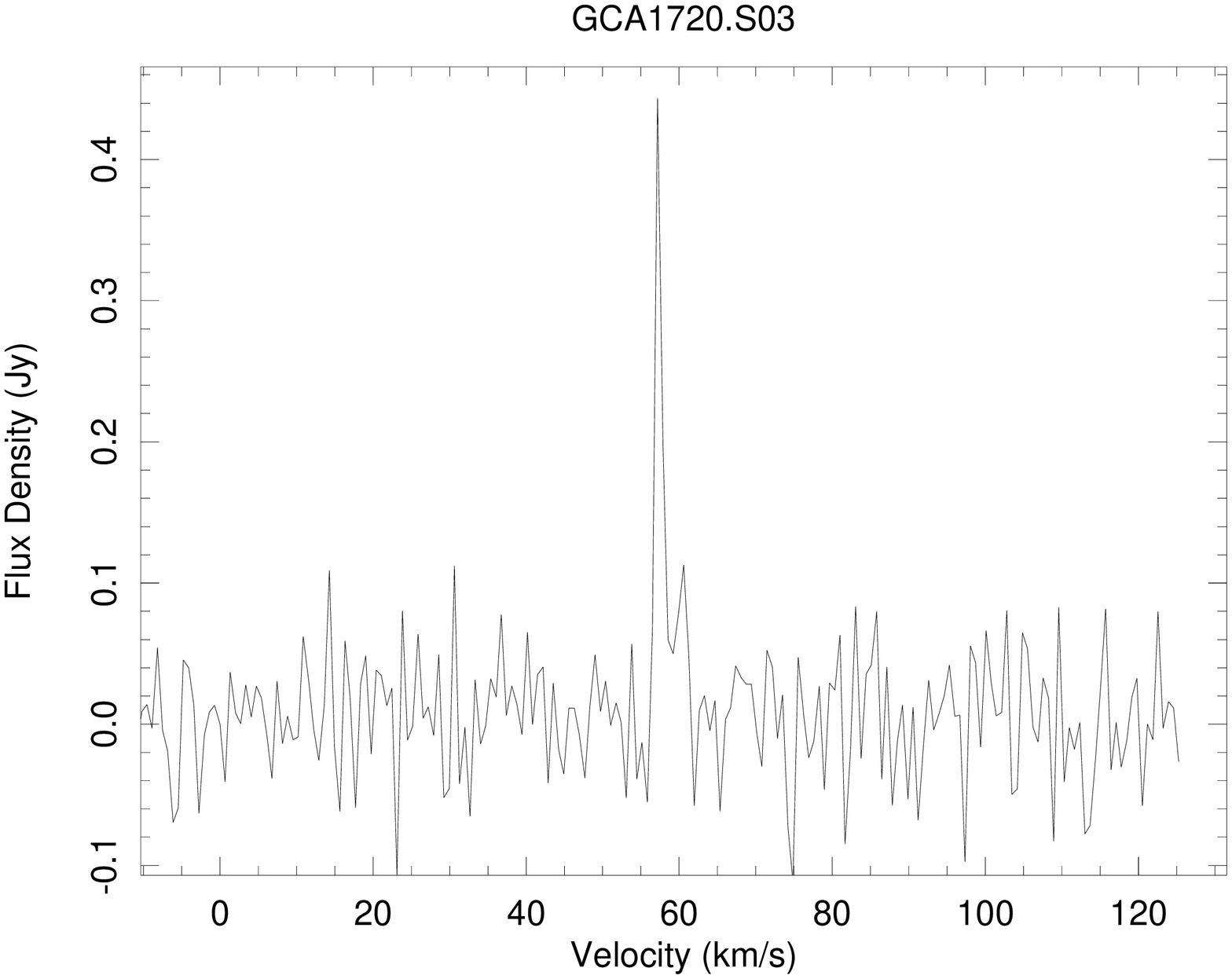}
\includegraphics[width=2.5in,angle=-90]{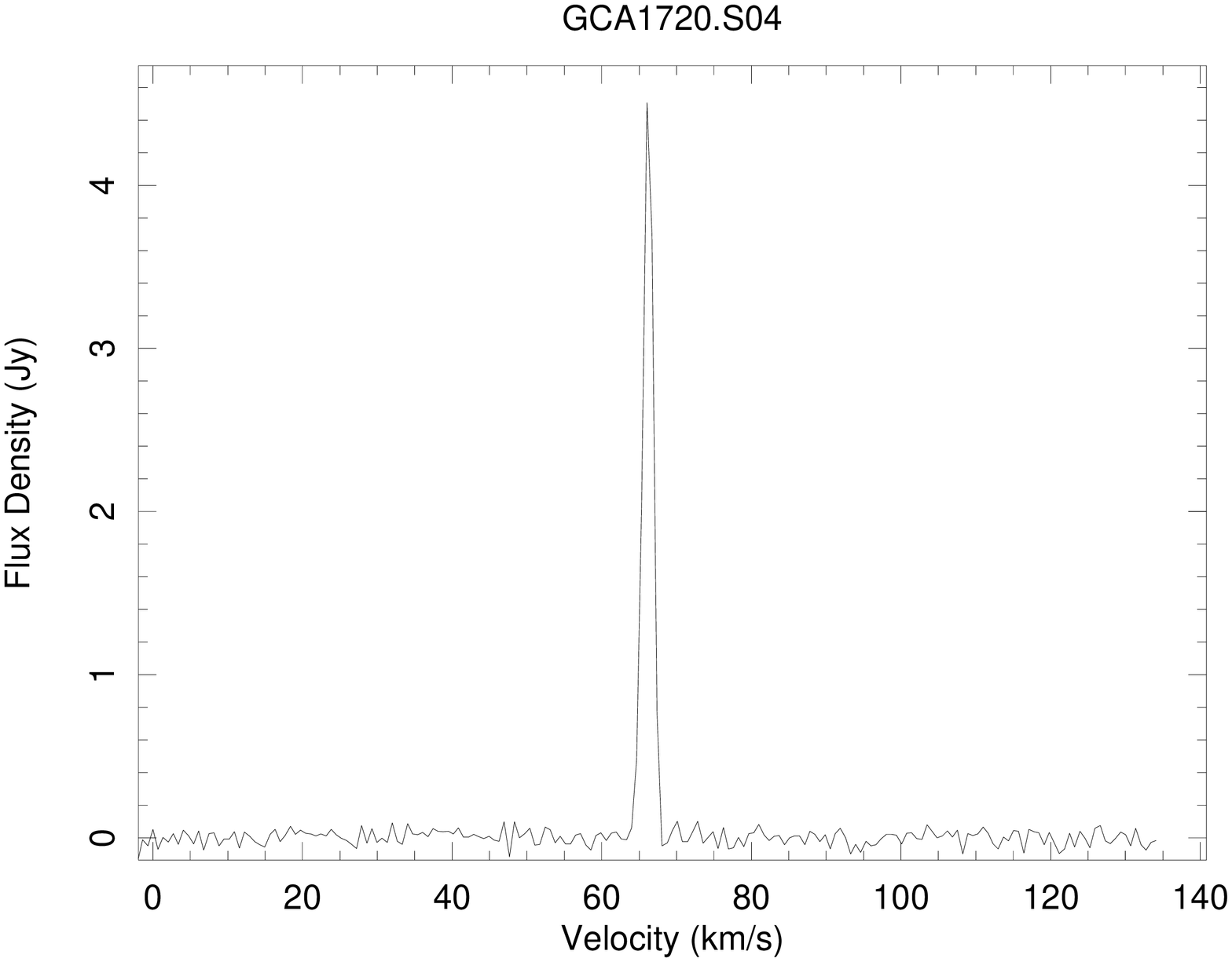}
}
\centerline{
\includegraphics[width=2.5in,angle=-90]{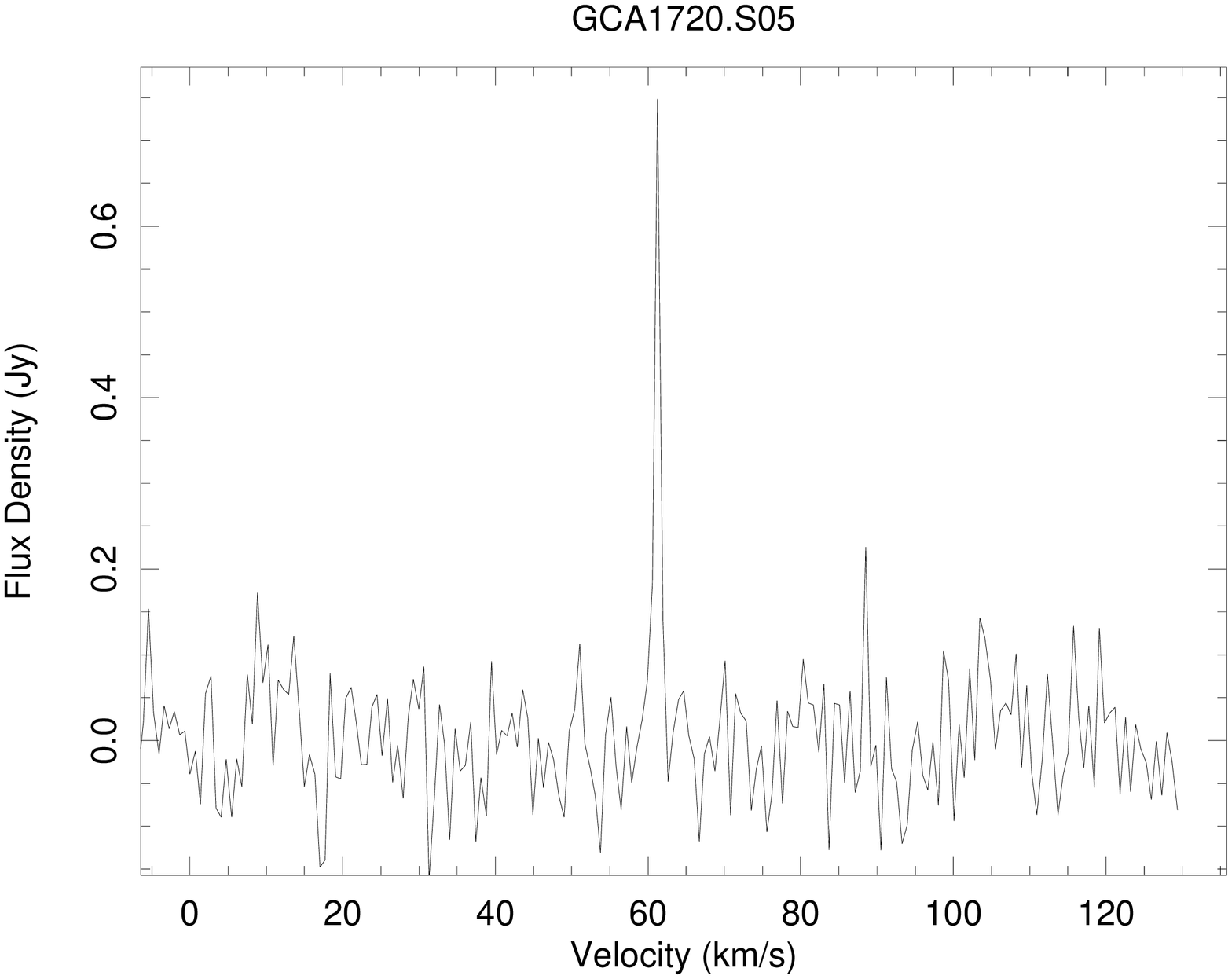}
}
\caption {1720 MHz OH maser spectra}
\label{OH1720Spectfig}
\end{figure*}

A list of the 1665 MHz OH sources is given in Table \ref{OH1665MaserCatalog};
sample spectra are shown in Figure \ref{OH1665Spectfig}.
The 1667 MHz OH sources are given in Table \ref{OH1667MaserCatalog} and
spectra are shown in Figure \ref{OH1667Spectfig}.
1720 MHz OH sources are listed in Table \ref{OH1720MaserCatalog} with spectra
displayed in Figure \ref{OH1720Spectfig}.



The brightness temperature limits on the OH sources given in Tables
\ref{OH1612MaserCatalog}--\ref{OH1720MaserCatalog} indicate a
nonthermal, i.e. maser, origin of these sources. 
Most of the 1612 MHz masers have a double horned spectrum
characteristic of evolved stars such as AGB stars although several are
very narrow, single peaked spectrum sources.
The 1665 MHz masers, as illustrated in Figure \ref{OH1665Spectfig}, have
velocity structure over a range of a few to several 10's of km
$s^{-1}$. 
The negative features in the 1665 MHz { spectra of GCA1665.09 and
GCA1665.10} are due to nearby strong masers like  { GCA1665.12}. 


The few 1667 MHz masers detected have a relatively simple velocity
structure with one or two components of width a few km $s^{-1}$.
The 1720 MHz masers detected are all very narrow in velocity.
The 1720 MHz maser  { GCA1720.05}, is associated with a non thermal continuum
source in Sgr B2(M) and is discussed in \cite{SgrB2OH} as possibly
associated with a SNR interacting with the Sgr B2 molecular cloud.

{ There have been numerous large scale OH surveys of the Galactic center
over the last three decades.  
Sensitive OH observations of the inner degree of the Galactic center
identified 1612 MHz masers associated with OHIR stars (Lundqvist et
al. 1992, 1995; Sjowermann et al. 1998).  
The detection of OH(1612 MHz) masers, as listed in Table 4, have all
been identified in past surveys. 
There have also been a large number of targeted OH surveys of the Sgr
A complex and Sgr B2 at 1720 and 1665, 1667 MHz (e.g., Caswell \&
Haynes 1983; Gaume \& Mutel 1987; Argon, Reid \& Menten 2000;
Yusef-Zadeh et al. 1996, 1999, 2016; Karlsson et al. 2003; Sjouwerman
et al.  1998, 2002; Caswell, Green and Phillips 2013). 
The OH (1720 MHz) masers identified in Table 7 are concentrated in the
Sgr A complex and Sgr B2, thus have been detected previously (Argon,
Reid \& Menten 2000; Yusef-Zadeh et al. 1999). 
Of the OH masers at 1665 and 1667 MHz, GCA1665.02, GCA1665.03,
GCA1665.05, GCA1667.01, GCA1667.04, have not been identified in
previous surveys.}

 \acknowledgments
{ We wish to acknowledge the numerous helpful comments made by the
anonymous reviewer which led to an improved paper.}
The National Radio Astronomy Observatory is a facility of the National
Science Foundation, operated under a cooperative agreement by Associated
Universities, Inc.
This work is partially supported by the grant
AST-1517246 from the National Science Foundation.
This research made use of data products from the Midcourse Space
Experiment. Processing of the data was funded by the Ballistic Missile
Defense Organization with additional support from NASA Office of Space
Science. This research has also made use of the NASA/ IPAC Infrared
Science Archive, which is operated by the Jet Propulsion Laboratory,
California Institute of Technology, under contract with the National
Aeronautics and Space Administration.

\clearpage


\clearpage

\begin{deluxetable}{rccccrr}
\tablecaption{36 GHz continuum Catalog\label{ContKaCat}}
\tablewidth{0pt}
\tablehead{
\colhead{Name} & \colhead{RA 2000} & \colhead{$\pm$} & \colhead{Dec 2000} &\colhead{$\pm$}& \colhead{Flux} & \colhead{$\pm$}\\ 
              &                   &  \colhead{$''$}    &                    & \colhead{$''$}   & \colhead{mJy beam$^{-1}$} & \colhead{mJy beam$^{-1}$} }
\startdata
GCA36.01 &  17 43 50.794 & 0.007  &-29 22 01.72 & 0.009  &    3.44 &  0.39 \\
GCA36.02 &  17 44 06.910 & 0.006  &-29 24 16.42 & 0.006  &    4.62 &  0.24 \\
GCA36.03 &  17 45 16.181 & 0.005  &-29 03 15.43 & 0.006  &    2.83 &  0.22 \\
GCA36.04 &  17 45 39.822 & 0.006  &-29 00 23.52 & 0.008  &   62.78 &  6.46 \\
GCA36.05 &  17 45 39.925 & 0.004  &-29 00 25.00 & 0.009  &   73.21 &  7.17 \\
GCA36.06 &  17 45 40.039 & 0.001  &-29 00 27.92 & 0.001  & 1160.05 &  8.20 \\
GCA36.07 &  17 45 40.094 & 0.006  &-29 00 31.93 & 0.008  &   58.37 &  7.28 \\
GCA36.08 &  17 45 58.318 & 0.001  &-28 53 32.82 & 0.001  &    7.71 &  0.23 \\
GCA36.09 &  17 46 04.468 & 0.006  &-28 34 21.47 & 0.010  &    4.21 &  0.52 \\
GCA36.10 &  17 46 10.536 & 0.001  &-28 55 50.03 & 0.001  &   16.02 &  0.33 \\
\enddata
\end{deluxetable}


\begin{deluxetable}{cccccrrrrrrr}
\rotate
\tablecaption{Sample Detected Methanol Masers\label{KaMaserCatalog}}
\tablewidth{0pt}
\tablehead{
\colhead{Name} & \colhead{RA 2000} & \colhead{$\pm$} & \colhead{Dec
  2000} &\colhead{$\pm$} & \colhead{G. long} & \colhead{G. lat} &
\colhead{Flux}&\colhead{$\pm$}  & \colhead{Vel}& \colhead{Width}&
\colhead{T$_{\rm B}$}\\ 
    &                &     \colhead{$''$} &                      &
\colhead{$''$}    & \colhead{$\circ$} &
\colhead{$\circ$}&\colhead{mJy}&\colhead{mJy} & \colhead{km s$^{-1}$}
& \colhead{km s$^{-1}$} & \colhead{10$^5$ K}}
\startdata
GCACH3OH.0001 & 17 43 51.894 & 0.004 & -29 25 16.61 & 0.008 &  -0.614 &   0.073 &      257 &     28 &   157.3 &     1.0 & $>$   0.34 \\
GCACH3OH.0002 & 17 43 51.947 & 0.008 & -29 24 53.59 & 0.013 &  -0.608 &   0.076 &      270 &     28 &   153.2 &     1.0 & $>$   0.36 \\
GCACH3OH.0003 & 17 43 52.060 & 0.008 & -29 24 49.26 & 0.007 &  -0.607 &   0.077 &      305 &     31 &   157.3 &     1.0 & $>$   0.41 \\
GCACH3OH.0004 & 17 43 52.328 & 0.004 & -29 24 16.16 & 0.006 &  -0.599 &   0.081 &      228 &     28 &   177.5 &     1.4 & $>$   0.30 \\
GCACH3OH.0005 & 17 43 53.234 & 0.003 & -29 24 18.79 & 0.006 &  -0.598 &   0.077 &      376 &     25 &   176.3 &     1.1 & $>$   0.50 \\
GCACH3OH.0006 & 17 43 53.267 & 0.005 & -29 24 19.24 & 0.007 &  -0.598 &   0.077 &      321 &     22 &   179.0 &     1.4 & $>$   0.43 \\
GCACH3OH.0007 & 17 43 53.289 & 0.004 & -29 24 15.71 & 0.006 &  -0.597 &   0.078 &      879 &     32 &   177.6 &     1.8 & $>$   1.17 \\
GCACH3OH.0008 & 17 43 53.290 & 0.005 & -29 24 17.05 & 0.009 &  -0.597 &   0.078 &      562 &     33 &   177.6 &     1.4 & $>$   0.75 \\
GCACH3OH.0009 & 17 43 53.911 & 0.002 & -29 24 10.93 & 0.002 &  -0.594 &   0.076 &      254 &     30 &   177.5 &     1.4 & $>$   0.34 \\
GCACH3OH.0010 & 17 43 54.893 & 0.008 & -29 24 06.48 & 0.012 &  -0.592 &   0.074 &      284 &     34 &   177.5 &     1.4 & $>$   0.38 \\
GCACH3OH.0011 & 17 43 55.075 & 0.002 & -29 24 34.19 & 0.002 &  -0.598 &   0.070 &      249 &     31 &   177.5 &     1.4 & $>$   0.33 \\
GCACH3OH.0012 & 17 43 55.242 & 0.005 & -29 24 15.75 & 0.009 &  -0.593 &   0.072 &      224 &     24 &   177.5 &     1.4 & $>$   0.30 \\
\enddata
\tablecomments{Table \ref{KaMaserCatalog} is published in its entirety in
the electronic edition of the Astrophysical Journal.\\
A portion is shown here for guidance regarding its form and content.}
\end{deluxetable}

\begin{deluxetable}{rccccrr}
\tablecaption{1.7 GHz continuum Catalog\label{ContLCat}}
\tablewidth{0pt}
\tablehead{
\colhead{Name} & \colhead{RA 2000} & \colhead{$\pm$} & \colhead{Dec 2000} &\colhead{$\pm$}& \colhead{Flux} & \colhead{$\pm$}\\ 
              &                   &  \colhead{$''$}    &                    & \colhead{$''$}   & \colhead{mJy beam$^{-1}$} & \colhead{mJy beam$^{-1}$} }
\startdata
GCA1.7.01 & 17 44 37.0792 & 0.11 & -28 57 09.164 & 0.11 &   75.76  &  3.709    \\
GCA1.7.02 & 17 45 01.4628 & 0.35 & -29 03 35.031 & 0.19 &   26.20  &  2.430    \\
GCA1.7.03 & 17 45 11.2478 & 0.20 & -29 06 00.345 & 0.22 &   32.41  &  2.266    \\
GCA1.7.04 & 17 45 11.3584 & 0.27 & -29 05 46.326 & 0.21 &   34.41  &  3.068    \\
GCA1.7.05 & 17 45 24.7825 & 0.33 & -28 53 16.822 & 0.26 &   43.91  &  4.351    \\
GCA1.7.06 & 17 45 40.0471 & 0.07 & -29 00 27.992 & 0.05 &  649.34  & 18.690    \\
GCA1.7.07 & 17 45 52.4887 & 0.06 & -28 20 25.724 & 0.10 &   70.43  &  2.801    \\
GCA1.7.08 & 17 46 05.8111 & 0.26 & -28 49 04.330 & 0.31 &   35.05  &  3.442    \\
GCA1.7.09 & 17 46 07.1865 & 0.32 & -28 45 57.140 & 0.16 &   39.12  &  3.357    \\
GCA1.7.10 & 17 46 07.2753 & 0.28 & -28 45 52.439 & 0.17 &   28.56  &  2.275    \\
GCA1.7.11 & 17 47 14.6736 & 0.14 & -28 26 55.193 & 0.34 &   63.41  &  5.939    \\
GCA1.7.12 & 17 47 14.7915 & 0.32 & -28 26 51.494 & 0.20 &   54.11  &  5.271    \\
GCA1.7.13 & 17 47 20.4756 & 0.08 & -28 23 45.033 & 0.07 &  276.95  & 11.153    \\
\enddata
\end{deluxetable}

\begin{deluxetable}{cccccrrrrrrlr}
\rotate
\tablecaption{Detected 1612 MHz OH Emission\label{OH1612MaserCatalog}}
\tablewidth{0pt}
\tablehead{
\colhead{Name} & \colhead{RA 2000} & \colhead{$\pm$} & \colhead{Dec 2000} &\colhead{$\pm$} & \colhead{G. long} & \colhead{G. lat} & \colhead{Flux}&\colhead{$\pm$}  & \colhead{Vel}& \colhead{Width} &\colhead{Type} &\colhead{T$_{\rm B}$}\\ 
    &                &     \colhead{"} &                      &
\colhead{"}    & \colhead{$\circ$} & \colhead{$\circ$}&\colhead{mJy}&\colhead{mJy} & \colhead{km s$^{-1}$} & \colhead{km s$^{-1}$} & & \colhead{10$^6$ K}}
\startdata
GCA1612.01 & 17 43 45.471 & 0.042 & -29 26 16.80 & 0.06 &  -0.640 &   0.084 &     3520 &     63 &  -198.4 &     2.1 & D  & $>$ 0.35\\
GCA1612.02 & 17 44 06.894 & 0.101 & -29 24 16.24 & 0.15 &  -0.571 &   0.036 &     1263 &     62 &     0.4 &     6.9 & S? & $>$ 0.13\\
GCA1612.03 & 17 44 34.972 & 0.102 & -29 04 35.54 & 0.15 &  -0.238 &   0.120 &     2249 &    107 &    10.0 &     2.4 & D  & $>$ 0.22\\
GCA1612.04 & 17 44 57.746 & 0.176 & -29 20 42.14 & 0.27 &  -0.424 &  -0.091 &      613 &     73 &   -72.4 &     1.8 & D  & $>$ 0.06\\
GCA1612.05 & 17 45 31.439 & 0.200 & -28 46 21.97 & 0.15 &   0.128 &   0.103 &      863 &     79 &   -40.6 &     3.1 & D? & $>$ 0.09\\
GCA1612.06 & 17 45 33.502 & 0.109 & -29 25 06.83 & 0.13 &  -0.419 &  -0.240 &     2819 &    120 &  -100.7 &     5.6 & D  & $>$ 0.28\\
GCA1612.07 & 17 45 38.620 & 0.194 & -28 59 45.24 & 0.21 &  -0.048 &  -0.035 &     1113 &     98 &    96.8 &     2.8 & S  & $>$ 0.11\\
GCA1612.08 & 17 45 40.178 & 0.122 & -28 59 47.49 & 0.15 &  -0.046 &  -0.041 &     2358 &    110 &    52.3 &     3.0 & D  & $>$ 0.24\\
GCA1612.09 & 17 45 40.482 & 0.109 & -29 05 02.94 & 0.06 &  -0.120 &  -0.087 &     3946 &    118 &   -32.1 &     3.3 & D  & $>$ 0.39\\
GCA1612.10 & 17 45 46.432 & 0.111 & -29 01 46.04 & 0.08 &  -0.062 &  -0.077 &     3285 &    107 &   -71.2 &     5.7 & D  & $>$ 0.33\\
GCA1612.11 & 17 45 49.416 & 0.246 & -28 58 48.76 & 0.18 &  -0.014 &  -0.061 &      815 &     89 &    33.8 &     3.8 & D  & $>$ 0.08\\
GCA1612.12 & 17 45 51.916 & 0.203 & -28 44 51.18 & 0.22 &   0.189 &   0.052 &      799 &     80 &    -8.3 &     2.1 & D? & $>$ 0.08\\
GCA1612.13 & 17 45 54.208 & 0.162 & -28 31 46.99 & 0.16 &   0.379 &   0.159 &     1112 &     77 &   126.1 &     3.3 & D  & $>$ 0.11\\
GCA1612.14 & 17 45 55.828 & 0.093 & -28 45 18.24 & 0.12 &   0.190 &   0.036 &     2442 &     96 &   148.2 &     5.9 & D  & $>$ 0.24\\
GCA1612.15 & 17 46 19.130 & 0.121 & -28 31 57.12 & 0.21 &   0.424 &   0.079 &      783 &     72 &   -21.4 &     3.3 & D  & $>$ 0.08\\
GCA1612.16 & 17 46 22.109 & 0.125 & -28 46 22.81 & 0.21 &   0.225 &  -0.055 &      951 &     87 &   -88.6 &     0.9 & D  & $>$ 0.10\\
GCA1612.17 & 17 46 32.135 & 0.061 & -28 41 04.16 & 0.09 &   0.319 &  -0.040 &     3128 &     85 &    92.6 &     4.5 & D  & $>$ 0.31\\
GCA1612.18 & 17 46 35.339 & 0.170 & -28 58 57.19 & 0.16 &   0.071 &  -0.205 &     1517 &    112 &   127.2 &     2.4 & D  & $>$ 0.15\\
GCA1612.19 & 17 46 39.039 & 0.161 & -28 28 06.75 & 0.10 &   0.517 &   0.050 &      980 &     69 &   185.3 &     1.4 & D  & $>$ 0.10\\
GCA1612.20 & 17 47 18.646 & 0.135 & -28 22 54.14 & 0.19 &   0.666 &  -0.029 &     1349 &     83 &    72.2 &     2.3 & D  & $>$ 0.13\\
GCA1612.21 & 17 47 20.146 & 0.211 & -28 23 05.10 & 0.13 &   0.667 &  -0.035 &     1657 &     97 &    65.2 &     8.9 & D  & $>$ 0.17\\
GCA1612.22 & 17 47 25.062 & 0.112 & -28 36 33.31 & 0.16 &   0.484 &  -0.167 &     1529 &     77 &   153.4 &     3.6 & D  & $>$ 0.15\\
GCA1612.23 & 17 47 25.434 & 0.044 & -28 23 36.38 & 0.06 &   0.669 &  -0.056 &     3363 &     92 &    68.4 &     1.2 & S  & $>$ 0.34\\
\enddata
\end{deluxetable}

\begin{deluxetable}{cccccrrrrrrr}
\rotate
\tablecaption{Detected 1665 MHz OH Emission\label{OH1665MaserCatalog}}
\tablewidth{0pt}
\tablehead{
\colhead{Name} & \colhead{RA 2000} & \colhead{$\pm$} & \colhead{Dec 2000} &\colhead{$\pm$} & \colhead{G. long} & \colhead{G. lat} & \colhead{Flux}&\colhead{$\pm$}  & \colhead{Vel}& \colhead{Width}& \colhead{T$_{\rm B}$}\\ 
    &                &     \colhead{$''$} &                      &
\colhead{$''$}    & \colhead{$\circ$} & \colhead{$\circ$}&\colhead{mJy}&\colhead{mJy} & \colhead{km s$^{-1}$} & \colhead{km s$^{-1}$}  & \colhead{10$^6$ K}}
\startdata
GCA1665.01 & 17 44 40.572 & 0.027 & -29 28 15.37 & 0.051 &  -0.564 &  -0.103 &     2957 &     54 &   -50.9 &     0.9 & $>$ 0.30 \\
GCA1665.02 & 17 45 39.085 & 0.064 & -29 23 29.71 & 0.096 &  -0.385 &  -0.243 &     2282 &     71 &    21.8 &     3.9 & $>$ 0.23 \\
GCA1665.03 & 17 46 21.399 & 0.014 & -28 35 38.99 & 0.026 &   0.376 &   0.040 &     6386 &     61 &    37.0 &     2.9 & $>$ 0.64 \\
GCA1665.04 & 17 46 28.696 & 0.156 & -29 20 29.25 & 0.228 &  -0.249 &  -0.371 &      565 &     61 &    -6.7 &     0.7 & $>$ 0.06 \\
GCA1665.05 & 17 46 37.438 & 0.122 & -28 37 29.91 & 0.398 &   0.380 &  -0.026 &     3760 &    448 &    68.5 &     0.7 & $>$ 0.38 \\
GCA1665.06 & 17 47 09.121 & 0.091 & -28 46 15.82 & 0.145 &   0.315 &  -0.201 &      913 &     46 &    27.0 &     2.1 & $>$ 0.09 \\
GCA1665.07 & 17 47 19.908 & 0.051 & -28 22 16.05 & 0.070 &   0.678 &  -0.027 &     4228 &    234 &    70.5 &     1.5 & $>$ 0.42 \\
GCA1665.08 & 17 47 20.000 & 0.131 & -28 22 55.73 & 0.275 &   0.669 &  -0.033 &      589 &     50 &    84.3 &     0.9 & $>$ 0.06 \\
GCA1665.09 & 17 47 20.042 & 0.033 & -28 22 40.75 & 0.063 &   0.672 &  -0.031 &     4139 &     94 &    55.2 &     2.4 & $>$ 0.41 \\
GCA1665.10 & 17 47 20.048 & 0.048 & -28 23 46.24 & 0.118 &   0.657 &  -0.041 &     1733 &     59 &    49.6 &     2.1 & $>$ 0.17 \\
GCA1665.11 & 17 47 20.107 & 0.037 & -28 23 05.12 & 0.040 &   0.667 &  -0.035 &    17827 &    201 &    61.6 &     2.5 & $>$ 1.78 \\
GCA1665.12 & 17 47 20.476 & 0.013 & -28 23 45.00 & 0.023 &   0.658 &  -0.042 &   143454 &   1236 &    69.0 &     1.9 & $>$14.33\\
GCA1665.13 & 17 47 45.067 & 0.014 & -28 44 28.08 & 0.017 &   0.409 &  -0.298 &     1154 &     85 &    68.5 &     0.9 & $>$ 0.12 \\
GCA1665.14 & 17 47 47.351 & 0.023 & -28 44 31.76 & 0.105 &   0.412 &  -0.305 &      471 &     54 &    69.4 &     1.7 & $>$ 0.05 \\
\enddata
\end{deluxetable}
\begin{deluxetable}{cccccrrrrrrrr}
\rotate
\tablecaption{Detected 1667 MHz OH Emission\label{OH1667MaserCatalog}}
\tablewidth{0pt}
\tablehead{
\colhead{Name} & \colhead{RA 2000} & \colhead{$\pm$} & \colhead{Dec 2000} &\colhead{$\pm$} & \colhead{G. long} & \colhead{G. lat} & \colhead{Flux}&\colhead{$\pm$}  & \colhead{Vel}& \colhead{Width} & \colhead{T$_{\rm B}$}\\ 
    &                &     \colhead{$''$} &                      &
\colhead{$''$}    & \colhead{$\circ$} & \colhead{$\circ$}&\colhead{mJy}&\colhead{mJy} & \colhead{km s$^{-1}$} & \colhead{km s$^{-1}$}  & \colhead{10$^6$ K}}
\startdata
GCA1667.01 & 17 44 51.304 & 0.081 & -29 24 54.56 & 0.097 &  -0.496 &  -0.107 &      418 &     37 &   -61.8 &     0.9 & $>$ 0.04 \\
GCA1667.02 & 17 45 33.515 & 0.107 & -29 25 06.92 & 0.170 &  -0.419 &  -0.240 &      507 &     47 &   -71.0 &     1.7 & $>$ 0.05 \\
GCA1667.03 & 17 45 38.605 & 0.077 & -28 59 45.26 & 0.077 &  -0.048 &  -0.035 &      429 &     50 &    69.4 &     0.9 & $>$ 0.04 \\
GCA1667.04 & 17 46 15.345 & 0.064 & -28 48 47.75 & 0.117 &   0.177 &  -0.055 &      505 &     33 &   -19.0 &     0.9 & $>$ 0.05 \\
GCA1667.05 & 17 47 20.402 & 0.508 & -28 23 03.88 & 0.166 &   0.667 &  -0.036 &      893 &     99 &    53.3 &     1.7 & $>$ 0.09 \\
\enddata
\end{deluxetable}
\begin{deluxetable}{cccccrrrrrrrr}
\rotate
\tablecaption{Detected 1720 MHz OH Emission\label{OH1720MaserCatalog}}
\tablewidth{0pt}
\tablehead{
\colhead{Name} & \colhead{RA 2000} & \colhead{$\pm$} & \colhead{Dec 2000} &\colhead{$\pm$} & \colhead{G. long} & \colhead{G. lat} & \colhead{Flux}&\colhead{$\pm$}  & \colhead{Vel}& \colhead{Width}& \colhead{T$_{\rm B}$}\\ 
    &                &     \colhead{$''$} &                      &
\colhead{$''$}    & \colhead{$\circ$} & \colhead{$\circ$}&\colhead{mJy}&\colhead{mJy} & \colhead{km s$^{-1}$} & \colhead{km s$^{-1}$} & \colhead{10$^6$ K}}
\startdata
GCA1720.01 & 17 45 38.800 & 0.106 & -28 59 42.57 & 0.169 &  -0.047 &  -0.036 &      765 &     57 &    44.3 &     0.9 & $>$ 0.08 \\
GCA1720.02 & 17 45 40.605 & 0.171 & -28 59 44.32 & 0.095 &  -0.044 &  -0.042 &     1303 &     61 &   133.1 &     1.0 & $>$ 0.13 \\
GCA1720.03 & 17 45 43.471 & 0.131 & -29 01 31.80 & 0.237 &  -0.064 &  -0.066 &      443 &     45 &    58.4 &     0.8 & $>$ 0.04 \\
GCA1720.04 & 17 45 44.335 & 0.055 & -29 01 18.98 & 0.031 &  -0.060 &  -0.067 &     4508 &     68 &    67.2 &     1.3 & $>$ 0.45 \\
GCA1720.05 & 17 47 20.029 & 0.148 & -28 23 12.15 & 0.220 &   0.665 &  -0.036 &      748 &     74 &    62.3 &     0.7 & $>$ 0.07 \\
\enddata
\end{deluxetable}


\begin{thebibliography}{}

\bibitem[Argon et al.(2000)]{2000ApJS..129..159A}
Argon, A.~L., Reid, M.~J., \& Menten, K.~M.\ 2000, \apjs, 129, 159

\bibitem[Avedisova (2002)]{avedisova}
Avedisova, V. S., 2002, { ARep}, 46, 193.

\bibitem[Bally et al.(1988)]{bally88}
Bally, J., Stark, A.~A., Wilson, R.~W., \& Henkel, C.\ 1988, \apj, 324, 223 

\bibitem[Caswell \& Haynes(1983)]{1983AuJPh..36..361C}
Caswell, J.~L., \& Haynes, R.~F.\ 1983, Australian Journal of Physics, 36, 361 

\bibitem[Caswell et al.(2013)]{2013MNRAS.431.1180C}
Caswell, J.~L., Green, J.~A., \& Phillips, C.~J.\ 2013, \mnras, 431, 1180  

\bibitem[Condon (1997)]{con97}
Condon, J.~J.~1997, PASP, 109, 166

\bibitem[Cotton (2008)]{Obit}
Cotton, W.~D. 2008, \pasp, 120, 439

\bibitem[Dahmen et al.(1997)]{dahmen97}
Dahmen, G., Huettemeister, S., Wilson, T.~L., et al.\ 1997, \aaps, 126  

\bibitem[Egan et al.(2003)]{MSX}
Egan, M. P., Price, S. D., Kramer, K. E., et al.\ 2003, Air Force Research Laboratory Technical Report AFRL-VS-TR-2003-1589 

\bibitem[Frail et al.(1994)]{frail94}
Frail, D.~A., Goss, W.~M., \& Slysh, V.~I.\ 1994, \apjl, 424, L111

\bibitem[Gaume \& Mutel(1987)]{1987ApJS...65..193G}
Gaume, R.~A., \& Mutel, R.~L.\ 1987, \apjs, 65, 193 

\bibitem[Hewitt et al.(2008)]{2008ApJ...683..189H}
Hewitt, J.~W., Yusef-Zadeh, F., \& Wardle, M.\ 2008, \apj, 683, 189

\bibitem[Huettemeister et al.(1993)]{huettemeister93}
Huettemeister, S., Wilson, T.~L., Bania, T.~M., \& Martin-Pintado, J.\ 1993, \aap, 280, 255 

\bibitem[Karlsson et al.(2003)]{2003A&A...403.1011K}
Karlsson, R., Sjouwerman, L.~O., Sandqvist, A., \& Whiteoak, J.~B.\ 2003, \aap, 403, 1011  

\bibitem[LaRosa et al.(2005)]{larosa05}
LaRosa, T.~N., Brogan, C.~L., Shore, S.~N., et al.\ 2005, \apjl, 626, L23 

\bibitem[Lindqvist et al.(1992)]{1992A&A...259..118L}
Lindqvist, M., Habing, H.~J., \& Winnberg, A.\ 1992, \aap, 259, 118  

\bibitem[Lindqvist et al.(1992)]{1992A&AS...92...43L}
Lindqvist, M., Winnberg, A., Habing, H.~J., \& Matthews, H.~E.\ 1992, \aaps, 92, 43  

\bibitem[Jones et al.(2012)]{jones12}
Jones, P.~A., Burton, M.~G., Cunningham, M.~R., et al.\ 2012, \mnras, 419, 2961 

\bibitem[Mehringer \& Menten (1997)]{mehringer97}
Mehringer, D.~M., \& Menten, K.~M.\ 1997, \apj, 474, 346  

\bibitem[Mills et al.(2015)]{mills15} 
Mills, E.~A.~C., Butterfield, N., Ludovici, D.~A., et al.\ 2015, \apj, 805, 72 

\bibitem[Oka et al.(1998)]{oka98}
Oka, T., Hasegawa, T., Sato, F., Tsuboi, M., \& Miyazaki, A.\ 1998, \apjs, 118, 455 

\bibitem[Oka et al.(2005)]{oka05}
Oka, T., Geballe, T.~R.,G oto, M., Usuda, T., \& McCall, B.~J.\ 2005, \apj, 632, 882

\bibitem[Roberts et al.(2007)]{roberts07}
Roberts, J.~F., Rawlings, J.~M.~C., Viti, S., \& Williams, D.~A.\ 2007, \mnras, 382, 733 

\bibitem[Sjouwerman et al.(1998)]{1998A&AS..128...35S}
Sjouwerman, L.~O., van Langevelde, H.~J., Winnberg, A., \& Habing, H.~J.\ 1998, \aaps, 128, 35  

\bibitem[Sjouwerman et al.(2002)]{2002A&A...391..967S}
Sjouwerman, L.~O., Lindqvist, M., van Langevelde, H.~J., \& Diamond, P.~J.\ 2002, \aap, 391, 967  

\bibitem[Sjouwerman et al.(2010)]{sjouwerman10}
Sjouwerman, L.~O., Pihlstr\"om, Y,~M. and Fish, V.~L. 2010, \apj\,  710, L111

\bibitem[Tsuboi et al.(1999)]{tsuboi99}
Tsuboi, M., Handa, T., \& Ukita, N.\ 1999, \apjs, 120, 1 


\bibitem[Voronkov et al.(2006)]{voronkov06}
Voronkov, M. A., Brooks, K. J., Sobolev, A. M. et al. 2006, MNRAS, 373, 411

\bibitem[Wardle (1999)]{wardle99}
Wardle, M.\ 1999, \apjl, 525, L101 

\bibitem[Yusef-Zadeh et al.(1996)]{1996ApJ...466L..25Y}
Yusef-Zadeh, F., Roberts, D.~A., Goss, W.~M., Frail, D.~A., \& Green, A.~J.\ 1996, \apjl, 466, L25  

\bibitem[Yusef-Zadeh et al.(1999)]{1999ApJ...512..230Y}
Yusef-Zadeh, F., Roberts, D.~A., Goss, W.~M., Frail, D.~A., \& Green, A.~J.\ 1999, \apj, 512, 230  

\bibitem[Yusef-Zadeh et al.(2013)]{Paper1}
Yusef-Zadeh, F., Cotton, W., Viti, S., Wardle, M., Royster, M. 2013 \apj, 764,  L19

\bibitem[Yusef-Zadeh et al.(2007)]{yusef-zadeh07}
Yusef-Zadeh, F., Wardle, M., \& Roy, S.\ 2007, \apj, 665, L123 

\bibitem[Yusef-Zadeh et al.(2013)]{yusef-zadeh13}
Yusef-Zadeh, F., Wardle, M., Lis, D., et al.\ 2013, Journal of Physical Chemistry A, 117, 9404 

\bibitem[Yusef-Zadeh et al.(2016)]{SgrB2OH}
Yusef-Zadeh, F. Cotton, W.,Wardle, M.,  Intema, H. 2016 \apj,  819, L35

\end{thebibliography}
\end{document}